%
%
%
%
%
%


\magnification=1200
\hsize=31pc
\vsize=55 truepc
\hfuzz=2pt
\vfuzz=4pt
\pretolerance=5000
\tolerance=5000
\parskip=0pt plus 1pt
\parindent=16pt
%

%
%
\font\fourteenrm=cmr10 scaled \magstep2
\font\fourteeni=cmmi10 scaled \magstep2
\font\fourteenbf=cmbx10 scaled \magstep2
\font\fourteenit=cmti10 scaled \magstep2
\font\fourteensy=cmsy10 scaled \magstep2

%
\font\large=cmbx10 scaled \magstep1

%

%

%

%
\font\eightrm=cmr8  
\font\eighti=cmmi8
\font\eightbf=cmbx8
\font\eightit=cmti8

\font\eightsy=cmsy8
\font\sixrm=cmr6
\font\sixi=cmmi6
\font\sixsy=cmsy6

%
\def\tenpoint{\def\rm{\fam0\tenrm}%
  \textfont0=\tenrm \scriptfont0=\sevenrm 
                      \scriptscriptfont0=\fiverm
  \textfont1=\teni  \scriptfont1=\seveni 
                      \scriptscriptfont1=\fivei
  \textfont2=\tensy \scriptfont2=\sevensy 
                      \scriptscriptfont2=\fivesy
  \textfont3=\tenex   \scriptfont3=\tenex 
                      \scriptscriptfont3=\tenex
  \textfont\itfam=\tenit  \def\it{\fam\itfam\tenit}%
  \textfont\slfam=\tensl  \def\sl{\fam\slfam\tensl}%
  \textfont\bffam=\tenbf  \scriptfont\bffam=\sevenbf
                            \scriptscriptfont\bffam=\fivebf
                            \def\bf{\fam\bffam\tenbf}%
  \normalbaselineskip=20 truept
  \setbox\strutbox=\hbox{\vrule height14pt depth6pt
width0pt}%
  \let\sc=\eightrm \normalbaselines\rm}
\def\eightpoint{\def\rm{\fam0\eightrm}%
  \textfont0=\eightrm \scriptfont0=\sixrm 
                      \scriptscriptfont0=\fiverm
  \textfont1=\eighti  \scriptfont1=\sixi
                      \scriptscriptfont1=\fivei
  \textfont2=\eightsy \scriptfont2=\sixsy
                      \scriptscriptfont2=\fivesy
  \textfont3=\tenex   \scriptfont3=\tenex
                      \scriptscriptfont3=\tenex
  \textfont\itfam=\eightit  \def\it{\fam\itfam\eightit}%
  \textfont\bffam=\eightbf  \def\bf{\fam\bffam\eightbf}%
  \normalbaselineskip=16 truept
  \setbox\strutbox=\hbox{\vrule height11pt depth5pt width0pt}}
\def\fourteenpoint{\def\rm{\fam0\fourteenrm}%
  \textfont0=\fourteenrm \scriptfont0=\tenrm 
                      \scriptscriptfont0=\eightrm
  \textfont1=\fourteeni  \scriptfont1=\teni 
                      \scriptscriptfont1=\eighti
  \textfont2=\fourteensy \scriptfont2=\tensy 
                      \scriptscriptfont2=\eightsy
  \textfont3=\tenex   \scriptfont3=\tenex 
                      \scriptscriptfont3=\tenex
  \textfont\itfam=\fourteenit  \def\it{\fam\itfam\fourteenit}%
  \textfont\bffam=\fourteenbf  \scriptfont\bffam=\tenbf
                             \scriptscriptfont\bffam=\eightbf
                             \def\bf{\fam\bffam\fourteenbf}%
  \normalbaselineskip=24 truept
  \setbox\strutbox=\hbox{\vrule height17pt depth7pt width0pt}%
  \let\sc=\tenrm \normalbaselines\rm}
\def\today{\number\day\ \ifcase\month\or
  January\or February\or March\or April\or May\or June\or
  July\or August\or September\or October\or November\or
December\fi
  \space \number\year}

%
\newcount\secno      
\newcount\subno      
\newcount\subsubno   
\newcount\appno      
\newcount\tableno    
\newcount\figureno   
%

%
\normalbaselineskip=20 truept
\baselineskip=20 truept

%
%
\def\title#1
   {\vglue1truein
   {\baselineskip=24 truept
    \pretolerance=10000
    \raggedright
    \noindent \fourteenpoint\bf #1\par}
    \vskip1truein minus36pt}
%

%
\def\author#1
  {{\pretolerance=10000
    \raggedright
    \noindent {\large #1}\par}}

%
\def\address#1
   {\bigskip
    \noindent \rm #1\par}

%
\def\shorttitle#1
   {\vfill
    \noindent \rm Short title: {\sl #1}\par
    \medskip}

%
\def\pacs#1
   {\noindent \rm PACS number(s): #1\par
    \medskip}

%
\def\jnl#1
   {\noindent \rm Submitted to: {\sl #1}\par
    \medskip}

%
\def\date
   {\noindent Date: \today\par
    \medskip}

%
\def\beginabstract
   {\vfill\eject
    \noindent {\bf Abstract. }\rm}

%
\def\keyword#1
   {\bigskip
    \noindent {\bf Keyword abstract: }\rm#1}

%
\def\endabstract
   {\par
    \vfill\eject}

%
%

%
\def\entry#1#2#3
   {\noindent
    \hangindent=20pt
    \hangafter=1
    \hbox to20pt{#1 \hss}#2\hfill #3\par}

%
\def\subentry#1#2#3
   {\noindent
    \hangindent=40pt
    \hangafter=1
    \hskip20pt\hbox to20pt{#1 \hss}#2\hfill #3\par}
\def\checkforsub{\futurelet\nexttok\decide}
\def\ssf{\relax}
\def\decide{\if\nexttok\ssf\let\endspace=\nospace
                \else\let\endspace=\extraspace\fi\endspace}
\def\nospace{\nobreak\par\nobreak}
%
%
\def\section#1{%
    \goodbreak
    \vskip50pt plus12pt minus12pt
    \nobreak
    \gdef\extraspace{\nobreak\bigskip\noindent\ignorespaces}%
    \noindent
    \subno=0 \subsubno=0 
    \global\advance\secno by 1
    \noindent {\bf \the\secno. #1}\par\checkforsub}

%
\def\subsection#1{%
     \goodbreak
     \vskip24pt plus12pt minus6pt
     \nobreak
     \gdef\extraspace{\nobreak\medskip\noindent\ignorespaces}%
     \noindent
     \subsubno=0
     \global\advance\subno by 1
     \noindent {\sl \the\secno.\the\subno. #1\par}\checkforsub}

%
\def\subsubsection#1{%
     \goodbreak
     \vskip20pt plus6pt minus6pt
     \nobreak\noindent
     \global\advance\subsubno by 1
     \noindent {\sl \the\secno.\the\subno.\the\subsubno. #1}\null. 
     \ignorespaces}

%
\def\appendix#1
   {\vskip0pt plus.1\vsize\penalty-250
    \vskip0pt plus-.1\vsize\vskip24pt plus12pt minus6pt
    \subno=0
    \global\advance\appno by 1
    \noindent {\bf Appendix \the\appno. #1\par}
    \bigskip
    \noindent}

%
\def\subappendix#1
   {\vskip-\lastskip
    \vskip36pt plus12pt minus12pt
    \bigbreak
    \global\advance\subno by 1
    \noindent {\sl \the\appno.\the\subno. #1\par}
    \nobreak
    \medskip
    \noindent}

%
\def\ack
   {\vskip-\lastskip
    \vskip36pt plus12pt minus12pt
    \bigbreak
    \noindent{\bf Acknowledgments\par}
    \nobreak
    \bigskip
    \noindent}


%

%
\def\tabcaption#1
   {\global\advance\tableno by 1
    \noindent {\bf Table \the\tableno.} \rm#1\par
    \bigskip}

%

%

%

%

%

%
\def\figcaption#1
   {\global\advance\figureno by 1
    \noindent {\bf Figure \the\figureno.} \rm#1\par
    \bigskip}

%
\def\references
     {\vfill\eject 
     {\noindent \bf References\par}
      \parindent=0pt
      \bigskip}

%

%
\def\refjl#1#2#3#4
   {\hangindent=16pt
    \hangafter=1
    \rm #1
   {\frenchspacing\sl #2
    \bf #3}
    #4\par}

%
\def\refbk#1#2#3
   {\hangindent=16pt
    \hangafter=1
    \rm #1
   {\frenchspacing\sl #2}
    #3\par}

%
\def\numrefjl#1#2#3#4#5
   {\parindent=40pt
    \hang
    \noindent
    \rm {\hbox to 30truept{\hss #1\quad}}#2
   {\frenchspacing\sl #3\/
    \bf #4}
    #5\par\parindent=16pt}

%
\def\numrefbk#1#2#3#4
   {\parindent=40pt
    \hang
    \noindent
    \rm {\hbox to 30truept{\hss #1\quad}}#2
   {\frenchspacing\sl #3\/}
    #4\par\parindent=16pt}

%

\def\ref#1{\noindent \hbox to 21pt{\hss 
#1\quad}\frenchspacing\ignorespaces}

%
\def\frac#1#2{{#1 \over #2}}

%

%

%


\chardef\ii="10

%

%

%

\catcode`\@=11
\def\vfootnote#1{\insert\footins\bgroup
    \interlinepenalty=\interfootnotelinepenalty
    \splittopskip=\ht\strutbox 
    \splitmaxdepth=\dp\strutbox \floatingpenalty=20000
    \leftskip=0pt \rightskip=0pt \spaceskip=0pt \xspaceskip=0pt
    \noindent\eightpoint\rm #1\ \ignorespaces\footstrut\futurelet\next\fo@t}

%
%
\def\eq(#1){\hfill\llap{(#1)}}
\catcode`\@=12
%
%



%
%
\def\CQG{Classical Quantum Grav.}





%
%

%
%

%
%

%

%

%

%
\def\gap{\;\lower3pt\hbox{$\buildrel > \over \sim$}\;}
%
%
\def\lap{\;\lower3pt\hbox{$\buildrel < \over \sim$}\;}
\def\tqs{\hbox to 25pt{\hfil}}


%
%
%
\def\LaTeX{L\kern-.26em \raise.6ex\hbox{\fiverm A}%
   \kern-.15em\TeX}%
\def\AmSTeX{%
{$\cal{A}$}\kern-.1667em\lower.5ex\hbox{%
 $\cal{M}$}\kern-.125em{$\cal{S}$}-\TeX}



\def\square{\hbox{\rlap{$\sqcap$}$\sqcup$}}

\title{Gravitational-wave tails of tails}
\author{Luc Blanchet}
\address{D\'epartement d'Astrophysique Relativiste et de Cosmologie
(UPR~176 du CNRS), Observatoire de Paris, 92195 Meudon Cedex, France}

\shorttitle{Gravitational-wave tails of tails}
\jnl{\CQG}
\date      

\beginabstract
The tails of gravitational waves are caused by scattering of linear waves
onto the space-time curvature generated by the total mass-energy of the
source.  Quite naturally, the tails of tails are caused by curvature
scattering of the tails of waves themselves.  The tails of tails are
associated with the cubic non-linear interaction between two mass
monopole moments and, dominantly, the mass quadrupole of the source.  In
this paper we determine the radiation field at large distances from the
source for this particular monopole-monopole-quadrupole interaction.  We
find that the tails of tails appear at the third post-Newtonian (3PN)
order beyond the usual quadrupole radiation.  Motivated by the need of
accurate templates to be used in the data analysis of future detectors
of gravitational waves, we compute the contribution of tails, and of
tails of tails, up to the 3.5PN order in the energy flux generated by
inspiraling compact binaries.
\endabstract

\section{Introduction}  

\subsection{Motivation and overview} 

Gravitational waves propagating through vacuum from their source to
infinity share all possible contributions associated with products of
multipole moments -- indeed, this is a consequence of the infinite
non-linearity of the field equations.

At the quadratic non-linear order, two multipole interactions play a
prominent role (and yield effects which are representative of that order). The
first interaction is that of the (mass-type) quadrupole moment
$M_{pq}(t)$, which dominates the radiation field for slowly-moving
sources, with itself.  See [1] and references therein for the
computation and discussion of this interaction (henceforth we shall
refer to [1] as paper I).  The second interaction is between $M_{pq}(t)$
and the static mass monopole moment $M$ of the source (Schwarzschild
mass, or, more precisely, ADM mass of the source).  Such interaction $M
\times M_{pq}(t)$ is physically due to the propagation of quadrupole
waves on the Schwarzschild background associated with $M$.  In
particular the scattering of waves onto the potential barrier of the
Schwarzschild metric produces the so-called tails, which are pieces of
the field depending on the parameters of the source at all instants from
$-\infty$ in the past up to the retarded time $t-r/c$.

The numerous works related or devoted to tails include some mathematical
investigations of curved space-time wave equations [2--9], several
investigations and constructions of post-Minkowskian expansions
[10--17], the studies of the linear perturbations of the Schwarzschild
metric by fields of various spins [18--24], some discussions of physical
properties of tails [25--31,53], and the development of accurate
wave-generation formalisms [54,55,32--35].  Not only the tails exist as
theoretical objects predicted by general relativity, but they should
exist in the future as real observed phenomena.  Indeed the presence of
tails in the gravitational-wave signals generated by inspiraling compact
binaries should be deciphered by the planned experiments VIRGO and LIGO
(see [36,37,30,31]).

In the present paper we develop further this subject by computing the
{\it cubic} interaction between the quadrupole $M_{pq}(t)$ and {\it two}
monopoles $M$. Physically this ``monopole-monopole-quadrupole''
interaction is responsible for the scattering of the linear quadrupole
waves $M_{pq}$ onto the second-order ($M^2$) potential barrier of the
Schwarzschild metric, and for the scattering of the quadratic tails
$M \times M_{pq}$ themselves onto the first-order ($M$) potential barrier.
The latter effect produces the so-called ``tails of tails'' of
gravitational waves, as they can pictorially be referred to.  In fact we
shall employ this crude appellation for the whole $M^2 \times M_{pq}$
interaction.

Like the quadratic tails, the cubic tails of tails could be computed using
black-hole perturbation techniques. In this paper, we rather employ the
particular post-Minkowskian approximation method proposed in [16], and
which is used to compute the quadratic metric $M_{pq} \times M_{rs}$ in
paper I.  In Section 2 we recall from previous work some relevant
results on linear and quadratic metrics, and we determine the cubic
source term in the field equations (in vacuum) corresponding to the
looked-for interaction $M^2 \times M_{pq}$.  Then we compute in Section
3 the (finite part of the) retarded integral of this cubic source term,
restricting ourselves to the leading order $1/r$ in the distance to the
source.  The complete radiation field for the $M^2 \times M_{pq}$
interaction is obtained, and discussed, in Section 4.

We find that the tails of tails carry a supplementary factor $1/c^6$ with
respect to the dominant quadrupole radiation, and therefore contribute
to the radiation field at the so-called third post-Newtonian (3PN)
order.  Such a high post-Newtonian order is {\it a priori} quite small
in absolute magnitude, but it is still relevant to the observations of
inspiraling compact binaries by VIRGO and LIGO [36--40,58].  Essentially
the post-Newtonian corrections in the field affect through gravitational
radiation reaction the evolution of the binary's orbital phase, and the
latter observable will be monitored very accurately in future detectors
thanks to the large number of observed periods of rotation.  In the
author's opinion, it is remarkable that such a cubically non-linear
effect as a tail of tail should be known in advance for comparison with
real observations -- and, therefore, should in principle be {\it
detected} at the same time.  In Section 5 we compute the tails of tails
at 3PN order occuring in the total energy flux generated by inspiraling
compact binaries (the energy flux is the crucial quantity to predict
because it yields via an energy balance argument the effects of
radiation reaction).  We compute also the contribution of quadratic
tails at the 3.5PN order (extending previous results at the 1.5PN and
2.5PN orders). Besides the tails of tails at 3PN in the energy flux of
binaries, there are also some ``instantaneous'' contributions, the computation
of which is in progress [56,57].

In the particular case where the mass of one body is very small as
compared with the other mass, the radiation field of compact binaries
has been computed analytically up to a very high post-Newtonian
approximation [48--51]:  notably up to the 4PN order [50] and, more
recently, 5.5PN order [51].  Our results in Section 5 are in perfect
agreement, wherever the comparison can be made, with the latter works.

The notation and conventions are essentially the same as in paper I (a
short summary is provided in [41]).  In order to reduce clutter we pose
$G=c=1$ when indicating the $G$'s and $c$'s is not essential.  For a
short review on the post-Minkowskian method we refer to Section 2 of
paper I.

\subsection{Notation for the field equations} 

Our basic field variable is $h^{\alpha\beta}= \sqrt{-g}
g^{\alpha\beta}-\eta^{\alpha\beta}$, with $g^{\alpha\beta}$ the
contravariant metric, $g$ the determinant of the covariant metric, and
$\eta^{\alpha\beta}$ the Minkowski metric in Minkowskian coordinates
(signature $-+++$).  Subject to the condition of harmonic coordinates,

$$
\partial_\beta h^{\alpha\beta}=0 \ , \eqno(1.1)
$$
the vacuum field equations read

$$
\square h^{\alpha \beta}=N^{\alpha \beta}(h,h)+M^{\alpha \beta}(h,h,h)
+O(h^4)  \ , \eqno(1.2)
$$
where $\square$ denotes the flat d'Alembertian operator, and where the
source term in the right side is made of an infinite sum of quadratic,
cubic, and so on, functionals of $h^{\alpha\beta}$ and its first and
second derivatives.  The quadratic and cubic terms are given by

$$\eqalignno{
 N^{\alpha\beta} (h,h) =&- h^{\mu\nu} \partial_\mu \partial_\nu
 h^{\alpha\beta} + {1\over 2} \partial^\alpha h_{\mu\nu} \partial^\beta
 h^{\mu\nu} - {1\over 4} \partial^\alpha h \partial^\beta h \cr
&-2 \partial^{(\alpha} h_{\mu\nu} \partial^\mu h^{\beta)\nu}
  +\partial_\nu h^{\alpha\mu} (\partial^\nu h^\beta_\mu + \partial_\mu
  h^{\beta\nu}) \cr
&+ \eta^{\alpha\beta} \left[ -{1\over 4}\partial_\rho h_{\mu\nu}
  \partial^\rho h^{\mu\nu} +{1\over 8}\partial_\mu h \partial^\mu h
  +{1\over 2}\partial_\mu h_{\nu\rho} \partial^\nu h^{\mu\rho}\right]\ ,
&(1.3)\cr}
$$
$$ \eqalignno{
M^{\alpha\beta} (h,h,h) &= - h^{\mu\nu} (\partial^\alpha h_{\mu\rho}
 \partial^\beta h^\rho_\nu + \partial_\rho h^\alpha_\mu \partial^\rho
h^\beta_\nu -\partial_\mu h^\alpha_\rho \partial_\nu h^{\beta\rho})\cr
& + h^{\alpha\beta} \left[ -{1\over 4}\partial_\rho h_{\mu\nu}
  \partial^\rho h^{\mu\nu} +{1\over 8}\partial_\mu h \partial^\mu h
+{1\over 2}\partial_\mu h_{\nu\rho} \partial^\nu h^{\mu\rho}\right]\cr
& +{1\over 2}h^{\mu\nu} \partial^{(\alpha} h_{\mu\nu} \partial^{\beta)}h
  +2 h^{\mu\nu}\partial_\rho h^{(\alpha}_\mu \partial^{\beta)} h^\rho_\nu
 \cr
& + h^{\mu(\alpha} \left( \partial^{\beta)} h_{\nu\rho}
 \partial_\mu h^{\nu\rho} -2\partial_\nu h^{\beta)}_\rho \partial_\mu
 h^{\nu\rho} -{1\over 2}\partial^{\beta)} h \partial_\mu h \right)\cr
& +\eta^{\alpha\beta}\left[ {1\over 8} h^{\mu\nu}\partial_\mu h\partial_\nu h
 - {1\over 4} h^{\mu\nu}\partial_\rho h_{\mu\nu} \partial^\rho h
 - {1\over 4} h^{\rho\sigma}\partial_\rho h_{\mu\nu} \partial_\sigma h^{\mu\nu}
\right.  \cr
& \qquad\left. - {1\over 2} h^{\rho\sigma} \partial_\mu h_{\rho\nu}
\partial^\nu h^\mu_\sigma + {1\over 2} h^{\rho\sigma} \partial_\mu
h^\nu_\rho \partial^\mu h_{\sigma\nu} \right] \ .
                   & (1.4)\cr } $$
The higher-order terms are even more complicated but will not be needed
in this paper. 
\section{The monopole-monopole-quadrupole source term} 
\subsection{The linear and quadratic metrics}  

We look for a solution of the equations (1.1)-(1.4) in the form of a
post-linear (or post-Minkowskian) expansion

$$  h^{\alpha\beta} = Gh^{\alpha\beta}_1 + G^2h^{\alpha\beta}_2
    + G^3h^{\alpha\beta}_3 + ... \ , \eqno(2.1) $$
where $G$ denotes the Newton constant.  The harmonic coordinate
condition (1.1) implies that all the coefficients of the $G^n$'s are
divergenceless.  On the other hand, the field equations (1.2) imply that
the coefficients of any $G^n$'s obey a d'Alembertian equation whose
source is known from the previous coefficients, i.e.  coefficients of
the $G^m$'s where $1 \leq m \leq n-1$ (see e.g.  Section 2 in paper I).

Our starting point is the linearized metric $h^{\alpha\beta}_1$ defined by
the equation (2.3) of paper I. This metric is in the form of a multipolar
series parametrized by symmetric and trace-free (STF) mass-type
multipole moments $M_L$ ($\ell \geq 0$) and current-type ones $S_L$
($\ell \geq 1$) [41].  These moments reduce, in the Newtonian limit $c
\to \infty$, to the usual Newtonian multipole moments [14].  As we are
ultimately interested only in the cubic interaction between two
monopoles $M$ and the quadrupole $M_{pq}$, we retain in the linearized
metric only the terms
involving $M$ and $M_{pq}$.  Accordingly we denote

$$ h^{\alpha\beta}_1 = h^{\alpha\beta}_{(M)} + h^{\alpha\beta}_{(M_{pq})}
     \ .\eqno (2.2) $$
The monopole term reads
$$\eqalignno{
  h^{00}_{(M)} &= - 4 r^{-1} M  \ , &(2.3a) \cr
  h^{0i}_{(M)} &= h^{ij}_{(M)} = 0  \ .&(2.3b) \cr } $$
This is simply the linearized piece of the Schwarzschild metric in harmonic
coordinates, for which only the $00$ component of our field variable is
non-zero. The quadrupole term in (2.2) is 

$$ \eqalignno{
  h^{00}_{(M_{pq})} &= - 2 \partial_{ab} \left[ r^{-1} M_{ab}
    (t - r) \right]  \ , &(2.4a) \cr
  h^{0i}_{(M_{pq})} &= 2 \partial_a \left[ r^{-1} M^{(1)}_{ai}
    (t - r) \right]  \ , &(2.4b) \cr
  h^{ij}_{(M_{pq})} &= - 2 r^{-1} M^{(2)}_{ij} (t - r) \ .&(2.4c) \cr } $$
We use the same notation as in (2.3) of paper I, which
gives the complete linearized metric including all multipole terms. In the
following we need rather the metric (2.4) in expanded form, where the
spatial derivatives acting on both $r^{-1}$ and $t-r$ are worked out. We have

$$ \eqalignno{
  h^{00}_{(M_{pq})} &= - 2 n_{ab} r^{-3} \left\{ 3 M_{ab} + 3
   r M^{(1)}_{ab} + r^2 M^{(2)}_{ab}
   \right\}  \ , &(2.5a) \cr
  h^{0i}_{(M_{pq})} &= - 2 n_a r^{-2} \left\{ M^{(1)}_{ai} +
   r M^{(2)}_{ai} \right\}  \ , &(2.5b) \cr
  h^{ij}_{(M_{pq})} &= - 2 r^{-1} M^{(2)}_{ij}  \ .&(2.5c) \cr } $$
Henceforth we generally do not indicate the dependence of the moments on $t-r$. 

Consider the quadratic metric $h^{\alpha\beta}_2$ generated by the linear
metric (2.2)-(2.5). It is clear that $h^{\alpha\beta}_2$ involves a term
proportional to $M^2$, the mixed term corresponding to the interaction
$M \times M_{pq}$, and the term corresponding to the self-interaction
of $M_{pq}$.  Thus,
$$ h^{\alpha\beta}_2 = h^{\alpha\beta}_{(M^2)} +
   h^{\alpha\beta}_{(MM_{pq})} + h^{\alpha\beta}_{(M_{pq} M_{rs})} \ .
   \eqno (2.6) $$
The first term is the quadratic piece of the Schwarzschild
metric in harmonic coordinates,
$$ \eqalignno{
  h^{00}_{(M^2)} &= - 7 r^{-2} M^2  \ , &(2.7a) \cr
  h^{0i}_{(M^2)} &= 0  \ , &(2.7b) \cr
  h^{ij}_{(M^2)} &= - n_{ij} r^{-2} M^2  \ ,&(2.7c) \cr } $$
with four-dimensional trace ($h\equiv\eta_{\alpha\beta} h^{\alpha\beta}$)

$$ h_{(M^2)} = 6 r^{-2} M^2  \ .\eqno (2.7d) $$
The term $h^{\alpha\beta}_{(MM_{pq})}$, which constitutes the dominant
non-static multipole interaction at the quadratic order, obeys a
d'Alembertian equation whose source is given by (1.3) where (2.3) and (2.5)
are inserted, and which can be written, with obvious notation, as
$N^{\alpha\beta}(h_{(M)},h_{(M_{pq})})+N^{\alpha\beta}(h_{(M_{pq})},h_{(M)})$.
The solution of this d'Alembertian equation, and, then, the complete
metric $h^{\alpha\beta}_{(MM_{pq})}$ itself, is obtained using the
present method.  The monopole-quadrupole metric is the analogue of the
2-2 metric in Bonnor's double series method [10].  We simply report the
result of its computation, which can be found in Appendix B of [32]:

$$\eqalignno{
M^{-1} h^{00}_{(MM_{pq})} &= n_{ab} r^{-4} \left\{ -21 M_{ab}
  -21 r M^{(1)}_{ab} + 7 r^2 M^{(2)}_{ab}
  + 10 r^3 M^{(3)}_{ab} \right\} \cr
  &+ 8 n_{ab} \int^{+\infty}_1 dx Q_2 (x) M^{(4)}_{ab}(t - rx)\ ,   
  &(2.8a) \cr
M^{-1} h^{0i}_{(MM_{pq})} &= n_{iab} r^{-3} \left\{
  -M^{(1)}_{ab}  - r M^{(2)}_{ab} - {1\over 3} r^2 M^{(3)}_{ab} \right\}  \cr
  &+ n_a r^{-3} \left\{ -5 M^{(1)}_{ai} - 5 r M^{(2)}_{ai} + {19\over 3} r^2
   M^{(3)}_{ai} \right\} \cr
   &+ 8 n_a \int^{+\infty}_1 dx Q_1 (x) M^{(4)}_{ai} (t - rx) \ , 
  &(2.8b) \cr
M^{-1} h^{ij}_{(MM_{pq})} &= n_{ijab} r^{-4} \left\{ -{15\over 2}
   M_{ab} - {15\over 2} r M^{(1)}_{ab} - 3 r^2 M^{(2)}_{ab} - {1\over 2} r^3
     M^{(3)}_{ab} \right\} \cr
  &+ \delta_{ij} n_{ab} r^{-4} \left\{ -{1\over 2} M_{ab}
  - {1\over 2} r M^{(1)}_{ab} - 2 r^2
  M^{(2)}_{ab} - {11\over 6} r^3 M^{(3)}_{ab}
  \right\} \cr
  &+ n_{a(i} r^{-4} \left\{ 6 M_{j)a} + 6 r M^{(1)}_{j)a}
+ 6 r^2 M^{(2)}_{j)a} + 4 r^3 M^{(3)}_{j)a} \right\} \cr
  &+ r^{-4} \left\{ - M_{ij} - rM^{(1)}_{ij} - 4
  r^2 M^{(2)}_{ij} - {11\over 3} r^3 M^{(3)}_{ij} \right\} \cr
   &+ 8 \int^{+\infty}_1 dx Q_0 (x) M^{(4)}_{ij} (t - rx) \ , 
  &(2.8c) \cr }$$
with four-dimensional trace 
$$\eqalignno{
 M^{-1} h_{(MM_{pq})} &= n_{ab} r^{-4} \left\{ 18 M_{ab} + 18 r
  M^{(1)}_{ab} - 10 r^2 M^{(2)}_{ab} -
  12 r^3 M^{(3)}_{ab} \right\} \cr
  &- 8 \ n_{ab} \int^{+\infty}_1 dx Q_2 (x) M^{(4)}_{ab}(t - rx)
    \ . &(2.8d) \cr }$$
The metric is composed of two types of terms, instantaneous terms
depending on the quadrupole moment at time $t-r$ only, and non-local (or
hereditary) integrals depending on all instants from $- \infty$ in the past
to $t-r$.  The usual tail effects are contained in the non-local
integrals of (2.8).  Note that the non-local integrals come exclusively
from the source terms whose radial dependence is $r^{-2}$ [see Section 3
of paper I and (3.1) below].  The integrals are expressed in (2.8) by
means of the Legendre function of the second kind $Q_\ell$ (with branch
cut from $-\infty$ to $1$), which is related to the Legendre polynomial
$P_\ell$ by

$$ \eqalignno{
 Q_{\ell} (x) &= {1 \over 2} \int^1_{-1} P_{\ell} (y) {dy\over x-y}
         &(2.9a) \cr
   &= {1\over 2} P_\ell (x) {\rm ln}
    \left({x+1 \over x-1} \right)- \sum^{ \ell}_{ j=1}
    {1 \over j} P_{\ell -j}(x) P_{j-1}(x) &(2.9b)\cr } $$
(the first of these relations being known as Neumann's formula for the
Legendre function, see e.g.  [52]).  See also (A.15) in Appendix A for
still another expression of the Legendre function.  For future reference
we quote here the expansion of $Q_\ell$ when $x \to 1$ (with $x > 1$),

$$ Q_\ell (x)= -{1\over 2} \ln \left( {x-1\over 2}\right) - \sum^\ell_{j=1}
   {1\over j} + O [(x-1) \ln (x-1)]  \ .\eqno(2.9c) $$
[On the other hand, recall that the Legendre function behaves like
$1/x^{\ell+1}$ when $x \to +\infty$.] Finally the term
$h^{\alpha\beta}_{(M_{pq} M_{rs})}$ in (2.6) is the
quadrupole-quadrupole metric whose computation has been the subject of
paper I (we do not need this term in the present paper).

Now the cubic metric $h^{\alpha\beta}_3$ is made out of all possible
interactions of three moments chosen (with repetition) among $M$ and
the quadrupole $M_{pq}$, and is therefore constituted of four terms,

$$h^{\alpha\beta}_3 = h^{\alpha\beta}_{(M^3)} + h^{\alpha\beta}_{(M^2
M_{pq})} + h^{\alpha\beta}_{(M M_{pq} M_{rs})} +
h^{\alpha\beta}_{(M_{pq} M_{rs} M_{tu})}  \ .\eqno (2.10) $$
Of these terms only the first one is known within the present approach
(before this paper): this is the cubic piece of the Schwarzschild metric
that we give here for completeness,
$$ \eqalignno{
  h^{00}_{(M^3)} &= - 8 r^{-3} M^3  \ , &(2.11a) \cr
  h^{0i}_{(M^3)} &= h^{ij}_{(M^3)} = 0  \ .&(2.11b) \cr } $$
The term $h^{\alpha\beta}_{(M^2 M_{pq})}$ is the
monopole-monopole-quadrupole metric which will be dealt with in the
present paper.  The two last terms, which involve at least the
interaction of two quadrupole moments, will be left undetermined for
the time being.

\subsection{Expression of the cubic source} 

The metric $h^{\alpha\beta}_{(M^2 M_{pq})}$ obeys the
harmonic-coordinates condition $\partial_\beta h^{\alpha\beta}_{(M^2
M_{pq})} = 0$ and a d'Alembertian equation whose source exhausts all
possibilities of generating the multipole interaction $M^2 \times
M_{pq}$ by means of linear and quadratic metrics.  From (1.2) one has
$$ \square h^{\alpha\beta}_{(M^2 M_{pq})} = \Lambda^{\alpha\beta}_{(M^2
   M_{pq})} \ ,  \eqno(2.12)  $$
where the source term $\Lambda^{\alpha\beta}_{(M^2 M_{pq})}$ is obtained
from $N^{\alpha\beta}$ and $M^{\alpha\beta}$ defined in (1.3) and (1.4) as

$$\eqalignno{
\Lambda^{\alpha\beta}_{(M^2 M_{pq})} &= N^{\alpha\beta} \left( h_{(M^2)},
  h_{(M_{pq})} \right) + N^{\alpha\beta} \left( h_{(M_{pq})}, h_{(M^2)}
  \right)  \cr
 &+ N^{\alpha\beta} \left( h_{(M)}, h_{(M M_{pq})} \right) +
  N^{\alpha\beta} \left( h_{(M M_{pq})}, h_{(M)} \right) \cr
 &+ M^{\alpha\beta} \left( h_{(M)}, h_{(M)}, h_{(M_{pq})} \right) +
  M^{\alpha\beta} \left( h_{(M)}, h_{(M_{pq})}, h_{(M)} \right) \cr
 &+ M^{\alpha\beta} \left( h_{(M_{pq})}, h_{(M)}, h_{(M)} \right)\ .
  &(2.13) \cr }$$
In the first and second lines a linear metric is coupled to a quadratic one,
while in the third and fourth lines three linear metrics are coupled together.  The metrics have been distributed with all possibilities on each slots of the non-linear sources (1.3) and (1.4).

Using the explicit formulas (2.3), (2.5), (2.7) and (2.8), one obtains
the cubic source term $\Lambda^{\alpha\beta}_{(M^2 M_{pq})}$ after a tedious
but straightforward computation. As the monopole-quadrupole metric (2.8)
involves some non-local integrals, so does $\Lambda^{\alpha\beta}_{(M^2
M_{pq})}$ which can thus be split into a local (instantaneous) part, say
$I^{\alpha\beta}_{(M^2 M_{pq})}$, and a non-local (tail) part,
$T^{\alpha\beta}_{(M^2 M_{p q })}$:

$$\Lambda^{\alpha\beta}_{(M^2 M_{pq})} = I^{\alpha\beta}_{(M^2 M_{pq})} +
T^{\alpha\beta}_{(M^2 M_{pq})}  \ .\eqno (2.14) $$
The result for the instantaneous part is 

$$\eqalignno{
M^{-2} I^{00}_{(M^2 M_{pq})} &= n_{ab} r^{-7} \biggl\{ -516
  M_{ab} - 516 r M^{(1)}_{ab}
  - 304 r^2 M^{(2)}_{ab} \cr
  &\qquad\qquad - 76 r^3 M^{(3)}_{ab} 
  + 108 r^4 M^{(4)}_{ab}
  + 40 r^5 M^{(5)}_{ab} \biggr\} \ , &(2.15a) \cr
M^{-2} I^{0i}_{(M^2 M_{pq})} &= \hat{n}_{iab} r^{-6} \biggl\{
  4 M^{(1)}_{ab} + 4 r M^{(2)}_{ab} - 16 r^2 M^{(3)}_{ab}
  + {4\over 3} r^3
  M^{(4)}_{ab} - {4\over 3} r^4 M^{(5)}_{ab}
  \biggr\} \cr
  &+ n_a r^{-6} \biggl\{
  -{372\over 5} M^{(1)}_{ai} - {372\over 5} r M^{(2)}_{ai}
  -{232\over 5} r^2   M^{(3)}_{ai} \cr
  &\qquad\qquad- {84\over 5} r^3 M^{(4)}_{ai}
 + {124\over 5} r^4 M^{(5)}_{ai} \biggr\} \ , &(2.15b) \cr
M^{-2} I^{ij}_{(M^2 M_{pq})} &= \hat{n}_{ijab} r^{-5} \biggl\{
  -190 M^{(2)}_{ab} - 118 r M^{(3)}_{ab} 
  - {92\over 3} r^2 M^{(4)}_{ab} - 2 r^3
   M^{(5)}_{ab} \biggr\} \cr
  &+ \delta_{ij} n_{ab} r^{-5} \biggl\{ {160\over 7}
  M^{(2)}_{ab} + {176\over 7} r M^{(3)}_{ab}
  - {596\over 21}
  r^2 M^{(4)}_{ab} - {160\over 21} r^3 M^{(5)}_{ab} \biggr\} \cr
  &+ \hat{n}_{a(i} r^{-5} \biggl\{ -{312\over 7}
  M^{(2)}_{j)a} - {248\over 7} r M^{(3)}_{j)a} 
  + {400\over 7}
  r^2 M^{(4)}_{j)a} + {104\over 7} r^3 M^{(5)}_{j)a} \biggr\} \cr
  &+ r^{-5} \biggl\{ -12 M^{(2)}_{ij} - {196\over 15}
  r M^{(3)}_{ij} - {56\over 5} r^2
  M^{(4)}_{ij} - {48\over 5} r^3 M^{(5)}_{ij}
  \biggr\} &(2.15c) \cr }$$
(we recall e.g. that $\hat{n}_{ijab}$ denotes the STF projection of
$n_{ijab} \equiv n_in_jn_an_b$ [41]).  The tail part is composed of sums of
products of local terms with derivatives of tail terms.  Using some
elementary properties of the Legendre function [namely $x Q_1(x)={2
\over 3} Q_2(x)+{1 \over 3} Q_0(x)$ and $x Q_2(x)={ 3 \over 5} Q_3(x)+{2
\over 5} Q_1(x)$], we obtain

$$\eqalignno{
M^{-2} T^{00}_{(M^2 M_{pq})} &= n_{ab} r^{-3}  
  \int^{+\infty}_1 dx \biggl\{ 96 Q_0 M^{(4)}_{ab} 
  + \left[ {272\over 5} Q_1 + {168\over 5} Q_3 \right]
  r M^{(5)}_{ab} \cr
 &\qquad\qquad\qquad\qquad+ 32 Q_2 r^2 M^{(6)}_{ab} \biggr\}\ , &(2.16a) \cr
M^{-2} T^{0i}_{(M^2 M_{pq})} &= \hat{n}_{iab} r^{-3} 
  \int^{+\infty}_1 dx \biggl\{ - 32 Q_1 M^{(4)}_{ab} 
  + \left[ -{32\over 3} Q_0 + {8\over 3} Q_2 \right] 
 r M^{(5)}_{ab} \biggr\} \cr
  &+ n_a r^{-3} \int^{+\infty}_1 dx \biggl\{ {96\over 5} Q_1 M^{(4)}_{ai}
  + \left[ {192\over 5} Q_0 + {112\over 5} Q_2 \right] r M^{(5)}_{ai} \cr
  &\qquad\qquad\qquad\qquad+ 32 Q_1 r^2 M^{(6)}_{ai} \biggr\}\ , &(2.16b) \cr
M^{-2} T^{ij}_{(M^2 M_{pq})} &= \hat{n}_{ijab} r^{-3} 
  \int^{+\infty}_1 dx \biggl\{ - 32 Q_2 M^{(4)}_{ab} 
  + \left[- {32\over 5} Q_1 - {48\over 5} Q_3 \right] 
  r M^{(5)}_{ab} \biggr\} \cr
  &+ \delta_{ij} n_{ab} r^{-3} \int^{+\infty}_1 dx
  \biggl\{ - {32\over 7} Q_2 M^{(4)}_{ab}  
  + \left[ - {208\over 7} Q_1 + {24\over 7} Q_3 \right] 
  r M^{(5)}_{ab} \biggr\} \cr
  &+ \hat{n}_{a(i} r^{-3} \int^{+\infty}_1 dx \biggl\{
  {96\over 7} Q_2 M^{(4)}_{j)a}
  + \left[ {2112\over 35} Q_1 - {192\over 35} Q_3 \right] 
  r M^{(5)}_{j)a} \biggr\} \cr  
  &+ r^{-3} \int^{+\infty}_1 dx \biggl\{ {32\over 5} Q_2 M^{(4)}_{ij}
  + \left[ {1536\over 75} Q_1 - {96\over 75} Q_3 \right] r M^{(5)}_{ij} \cr
 &\qquad\qquad\qquad\qquad+ 32 Q_0 r^2 M^{(6)}_{ij} \biggr\}\ , 
  &(2.16c) \cr }$$
(where the Legendre functions are computed at $x$ and the moments at $t-rx$).
At this stage we have a good check that the computation is going well.
Indeed $\Lambda^{\alpha\beta}_{(M^2 M_{pq})}$ is the source of the third-order field equations in harmonic coordinates and therefore should be divergenceless. The divergence of the tail piece (2.16) is computed using the Legendre equation $(1-x^2) {Q''}_
\ell(x)-2x {Q'}_\ell(x)+\ell(\ell+1)Q_\ell(x)=\delta_+(x-1)$, where the
distribution $\delta_+$ is defined by $\int^{+\infty}_1 dx
\delta_+(x-1)\phi(x)=\phi(1)$ for any test function $\phi$. We find
that all the tail integrals disappear, and get

$$ \eqalignno{
M^{-2} \partial_\beta T^{0\beta}_{(M^2 M_{pq})} &= n_{ab}
  r^{-4} \Bigl\{ -48 M^{(4)}_{ab} - 48 r M^{(5)}_{ab} - 16
  r^2 M^{(6)}_{ab} \Bigr\}  \ , &(2.17a) \cr
M^{-2} \partial_\beta T^{i\beta}_{(M^2 M_{pq})} &= n_{iab}
  r^{-4} \Bigl\{ -8 M^{(4)}_{ab} - 16 r M^{(5)}_{ab} \Bigr\}
  - 32 n_a r^{-2} M^{(6)}_{ai}\ .\qquad &(2.17b) \cr }$$
On the other hand the divergence of the instantaneous piece is computed
directly from (2.15).  We readily obtain that it cancels exactly (2.17)
so that the required condition

$$\partial_\beta \Lambda^{\alpha\beta}_{(M^2 M_{pq})} = 0  \eqno (2.18) $$
is fulfilled. 

\section{The retarded integral of the source term} 

The previous check being done we confidently tackle the difficult part
of the analysis, namely to find the inversion of the d'Alembert equation
with source term given by (2.14)-(2.16).  We shall limit ourselves to
the computation of the metric at large distances from the source ($r
\to \infty$ with $t-r=$const), keeping only the dominant $1/r$ term at
infinity (possibly multiplied by some powers of $\ln r$).  This is
sufficient in view of applications to astrophysical sources of
gravitational radiation.

\subsection{Integrating the instantaneous terms} 

The general form of the terms composing (2.15) is $\hat{n}_L r^{-k}
F(t-r)$, where $\hat{n}_L$ is equivalent to a spherical harmonics of
order $\ell$, and where the radial dependence is such that $k \geq 2$.
So let us review the formulas required to compute the $1/r$ (and $\ln
r/r$) terms at infinity of the retarded integral of any source term
$\hat{n}_L r^{-k} F(t-r)$.  Recall that in general the retarded integral
cannot be applied directly on $\hat{n}_L r^{-k} F(t-r)$ because of the
singular behaviour when $r \to 0$.  We follow the procedure proposed in
[16] to obtain a particular retarded solution (also singular when $r \to
0$) of the wave equation.  It consists of multiplying first the source
term by a factor $(r/r_0)^B$, where $B$ is a complex number and $r _0$ a
constant length scale, thereby defining a $B$-dependent fictituous
source which, for large values of the real part of $B$, is regular,
and in fact tends to zero, when $r \to 0$ (i.e.  the singularity at
$r=0$ is killed).  For large values of Real($B$) one is thus allowed to
apply the retarded integral on the fictituous source (there is no
problem at the bound $r \to \infty$ of the integral because the moments
are assumed to be constant in the remote past).  In this way one defines
a function of $B $, {\it a priori} only for Real($B$) large enough, but
the point is that this function is extendible by analytic continuation
to all complex values of $B$, except at some integer values
(including in general the value of interest $B=0$), where it admits a
Laurent expansion with some poles.  Now it has been shown [16] that near
the value $B=0$ the finite part of the Laurent expansion {\it is} a
particular solution of the d'Alembertian equation we wanted to solve. We
call the operator giving this solution the finite part of the retarded
integral, and denote it by ${\rm FP}_{B=0} \square^{-1}_R (r/r_0)^B$, or
more simply by ${\rm FP} \square^{-1}_R$.  The finite part procedure is
especially convenient when doing practical computations.  See Appendix A
in paper I for a compendium of formulas, obtained with this procedure,
enabling the computation of the quadratic non-linearities.  More
complicated formulas to compute the cubic non-linearities are reported
in Appendix A of the present paper.

For a source term of the type $\hat{n}_L r^{-k} F(t-r)$ three cases must
be distinguished (as we have already $k \geq 2$):  $k = 2$, $3 \leq k
\leq \ell+2$, and $\ell+3 \leq k$ (see Appendix A of paper I).  The case
$k = 2$ corresponds to a retarded integral which is convergent (so the
finite part at $B=0$ is unnecessary), and given by a non-local integral
admitting an expansion when $r \to \infty$, $t-r=$ const in powers of
$1/r$ with a logarithm of $r$.  The formula, already used in (2.8) to
express the tail integrals, reads

$$ {\hbox{\rlap{$\sqcap$}$\sqcup$}}^{-1}_R \left[ \hat n_L r^{-2}
  F (t- r) \right] = - \hat n_L \int^{+\infty}_1 dx Q_{\ell}(x)
F(t-rx) \ , \eqno(3.1) $$
where $Q_\ell$ is the Legendre function (2.9).  The leading term at
infinity is obtained by using the expansion of $Q_\ell$ as given by
(2.9c).  We obtain [32]

$$ \eqalignno{
  {\hbox{\rlap{$\sqcap$}$\sqcup$}}^{-1}_R \left[ \hat n_L  r^{-2} F
   (t-r) \right]
  &= {\hat n_L\over 2r} \int^{+\infty}_0 d\tau
  F (t-r -\tau)
    \biggl[\ln \left( {\tau \over 2r} \right)
   + \sum^{\ell}_{j=1}{2 \over j} \biggr] \cr &+ O \left({\ln r \over r^2}
\right)\ . &(3.2)  \cr   } $$
In the case $3 \leq k \leq \ell+2$ (thus $\ell \geq 1$), we find that the
$B$-dependent integral is finite (no pole at $B=0$), and is given by
a simple local expression, without logarithms.  The $1/r$ term at
infinity is

$$ \eqalignno{
 \square^{-1}_R \bigl[(r/r_0)^B \hat n_L r^{-k}
 F(t-r) \bigr]_{|_{B=0}} =&  
  - {2^{k-3} (k-3)!(\ell+2-k)!\over (\ell+k-  2)!}
  {{\hat n_L}\over r} F^{(k-3)} (t-r)\cr
 &+ O\left( {1\over r^2} \right) \ .& (3.3) \cr  } $$
In the last case $k \geq \ell +3$, the $B$-dependent retarded integral
admits truly a polar part, and its finite part is given like in (3.2) by
a non-local integral, but now the expansion at infinity involves no
logarithms of $r$ (instead it involves the logarithm of the constant
$r_0$).  We have, for the $1/r$ term,

$$ \eqalignno{
  \hbox{FP} \square^{-1}_R \bigl[
   \hat n_L r^{-k} &F(t-r) \bigr]
  ={(-)^{k+\ell}2^{k-3}(k-3)!\over (k+\ell-2)!(k-\ell-3)!}{\hat n_L\over r}\cr
  & \times \int^{+\infty}_0 d\tau F^{(k-2)}
  (t-r -\tau) \biggl[ \ln \left( {\tau\over 2r_0}\right)
 +\sum^{k-\ell-3}_{j=1} {1\over j}
 +\sum^{k+\ell-2}_{j=k-2} {1\over j} \biggr] \cr
 &+ O \left( {1\over r^2} \right) \ .& (3.4) \cr } $$

With the formulas (3.2)-(3.4) it is straightforward to obtain the $1/r$
and $\ln r/r$ terms in the finite part of the retarded integral [in
short ${\rm FP} \square^{-1}_R$] of (2.15):

$$\eqalignno{
{\rm FP} \square^{-1}_R &\left[ M^{-2} I^{00}_{(M^2 M_{pq})} \right] =
  {n_{ab}\over r} \int^{+\infty}_0
  d\tau M^{(5)}_{ab} \left\{ 20 \ln \left( {\tau\over 2r} \right) 
  + {116\over 21}
  \ln \left( {\tau\over 2r_0} \right) + {106054\over 2205}
  \right\} \cr
  &+ O \left( {\ln r\over r^2} \right) \ , &(3.5a) \cr
{\rm FP} \square^{-1}_R &\left[ M^{-2} I^{0i}_{(M^2 M_{pq})}
  \right] = {\hat{n}_{iab}\over r}
  \int^{+\infty}_0 d\tau M^{(5)}_{ab} 
 \left\{ -{2\over 3}\ln \left( {\tau\over 2r} \right) - {4\over 105}
 \ln \left( {\tau\over 2r_0} \right) - {26044\over 11025}  \right\} \cr
  &+ {n_a\over r} \int^{+\infty}_0 d\tau M^{(5)}_{ia}
  \left\{ {62\over 5} \ln \left( {\tau\over 2r} \right) +
  {416\over 75} \ln \left( {\tau\over 2r_0} \right) + {40318\over
  1125} \right\} \cr
  &+ O \left( {\ln r\over r^2} \right) \ , &(3.5b) \cr
  {\rm FP} \square^{-1}_R &\left[ M^{-2} I^{ij}_{(M^2 M_{pq})}
  \right] = {\hat{n}_{ijab}\over r} \int^{+\infty}_0 d\tau M^{(5)}_{ab}
  \left\{ -\ln \left( {\tau\over 2r} \right) - {176\over 105} \right\} \cr
  &+ {\delta_{ij} n_{ab}\over r} \int^{+\infty}_0 d\tau   M^{(5)}_{ab}
  \left\{ - {80\over 21} \ln \left( {\tau\over 2r} \right) -
  {32\over 21} \ln \left( {\tau\over 2r_0} \right) - {3146\over 315}
  \right\} \cr
  &+ {\hat{n}_{a(i}\over r} \int^{+\infty}_0 d\tau   M^{(5)}_{j)a} 
  \left\{ {52\over 7} \ln \left( {\tau\over 2r} \right) +
  {104\over 35} \ln \left( {\tau\over 2r_0} \right) + {9472\over 525}
  \right\} \cr
  &+ {1\over r} \int^{+\infty}_0 d\tau M^{(5)}_{ij}
  \left\{ -{24\over 5} \ln \left( {\tau\over 2r} \right) +
 {92\over 15} \ln \left({\tau\over 2r_0}\right) + {94\over 15} \right\} \cr
  &+ O \left( {\ln r\over r^2} \right) \ , &(3.5c) \cr }
$$
where the moments are evaluated at time $t-r-\tau$.  Recall that in the
case of the quadratic non-linearities, the non-local integrals come only
from the source terms having $k=2$ (see Section 3 of paper I).  This is
not true in the case of the cubic (and higher) non-linearities:  besides
the integrals generated by the terms $k=2$, there are some non-local
integrals generated by terms having $k \geq \ell +3$.  These integrals,
given by (3.4), depend on the constant length scale $r_0$.  Thus the
particular solution which is picked up by the finite part procedure
depends on the constant $r_0$ used in its definition.
This is quite normal, and, of course, not a problem -- one needs only to
be consistent in using this particular solution, notably when relating
the multipole moments which parametrize it to the source variables.

\subsection{Integrating the tail terms}   

The effects which are physically associated with tails of tails come from
the retarded integration of the tail part of the cubic source, that is
$T^{\alpha\beta}_{(M^2 M_{pq})}$ given by (2.16).  The problem is to
find the (finite part of the) retarded integral of a source term
involving a non-local integral with some Legendre function $Q_m$, namely

$$_{k,m}\Psi_L \equiv {\rm FP}_{B=0} \square^{-1}_R \left[(r/r_0)^B
\hat{n}_L r^{-k} \int^{+\infty}_1 dx Q_m(x) F (t - rx)
\right]  \ . \eqno (3.6) $$
We study only the needed cases which correspond to $k \geq 1$ (note that
with this notation the actual radial dependence of the source when $r
\to \infty$, $t-r=$ const is $1/r^{k+1}$).  Of course the problem is
more complicated than in Section 3.1 because the source term at a given
value of $t-r$ is a more complicated function of $r$.

The detailed computation of $_{k,m}\Psi_L$ is relegated in Appendix A.
Here we report the main results.  The case $k=1$ is the most interesting
because it leads to qualitatively new results with respect to the
quadratic non-linear order.  When $k=1$ we find the solution in closed
analytic form,

$$\eqalignno{
  _{1,m}\Psi_L = &~\hat{n}_L \int^{+\infty}_1 dy F^{(-1)} (t - ry) \cr
  &\times \left\{ Q_\ell (y) \int^y_1 dx Q_m (x) {dP_\ell\over dx} (x)
  + P_\ell (y)   \int^{+\infty}_y dx Q_m (x)
  {dQ_\ell\over dx} (x) \right\}  \ ,\quad &(3.7) \cr }$$
where the time anti-derivative is defined by $F^{(-1)}(t) =
\int^t_{-\infty}dt'F(t')$ (all the functions involved are zero in the remote
past by our assumption of initial stationarity). The solution (3.7) has
been obtained thanks in particular to the mathematical formula (A.5) in
Appendix A.  To leading order when $r \to \infty$, $t-r=$ const the
second integral in the brackets dominates the first one.  Inspection of
(2.16) shows that when $k=1$ we need only the case $m =\ell$.  Because
the second integral in (3.7) can be explicitly worked out in this case,
we have a good simplification, since notably the dominant term at
infinity follows simply from the expansion (2.9c) of the Legendre
function.  We find a non-local expression with $\ln r$ and $\ln^2r$
terms:

$$\eqalignno{
 _{1,\ell}\Psi_L = &-{\hat{n}_L\over 8r} \int^{+\infty}_0 d\tau F^{(-1)}
  (t-r - \tau) \cr
  &\times \biggl[ \ln^2 \left( {\tau\over 2r} \right) + 4 \biggl(
  \sum^\ell_{j=1} {1\over j} \biggr) \ln \left( {\tau\over 2r} \right) + 4
  \biggl( \sum^\ell_{j=1} {1\over j} \biggr)^2 ~\biggr] +
  o\left(r^{\varepsilon-2}\right) \ .&(3.8) \cr }$$
For the remainder we use the notation $o(r^{\varepsilon-2})$ with a
small $o$-symbol to mean that the product of this remainder with the
factor $r^{2-\varepsilon}$, where $\varepsilon$ is such that $0 <
\varepsilon \ll 1$, tends to zero when $r \to \infty$.  This notation
is simply to account for the presence of powers of logarithms in the
expansion at infinity.

In the case $2 \leq k \leq \ell+2$, the result is simpler (see Appendix
A) in the sense that it is given by a local expression to order $1/r$,

$$_{k,m}\Psi_L = -~{}_{k,m}\alpha_\ell ~{\hat{n}_L\over r} F^{(k-3)}
(t-r) + o \left( {r^{\varepsilon-2}} \right)  \ . \eqno (3.9)  $$
However, the coefficient is still complicated: 

$$_{k,m}\alpha_\ell = \int^{+\infty}_1 dx Q_m (x) \int^{+\infty}_x dz
{(z-x)^{k-3}\over (k-3)!} Q_\ell (z) \ . \eqno (3.10) $$ 
The numerical values of the $_{k,m}\alpha_\ell$'s for general $k$,
$\ell$, and $m$ are computed in (A.17)-(A.18).  Here we need only some
values corresponding to $k=2$ and $k=3$ [see (2.16)].  These are
presented in Tables 1 and 2 respectively.

In the last case $k \geq \ell+3$, we obtain again a non-local integral,
but with simply a $1/r$ term without $\ln r$ (instead, there is a $\ln
r_0$),

$$
_{k,m}\Psi_L = -{\hat{n}_L\over r} \int^{+\infty}_0 d\tau
F^{(k-2)} (t-r - \tau)\left[ {}_{k,m}\beta_\ell \ln
\left( {\tau\over 2r_0} \right) +
{}_{k,m}\gamma_\ell \right] + o \left( {r^{\varepsilon-2}} \right)
 \ . \eqno(3.11)$$
The coefficients ${}_{k,m}\beta_\ell$ and ${}_{k,m}\gamma_\ell$ are also
rather involved,

$$ \eqalignno{
 _{k,m}\beta_\ell &= {1\over 2} \int^{+\infty}_1 dx Q_m (x) \int^1_{-1} dz
  {(z-x)^{k-3}\over (k-3)!} P_\ell (z)  \ , &(3.12a) \cr
 _{k,m}\gamma_\ell &= {1\over 2} \int^{+\infty}_1 dx Q_m (x) \int^1_{-1} dz
  {(z-x)^{k-3}\over (k-3)!} P_\ell (z)
  \biggl[ - \ln \left( {x-z\over 2} \right)
 + \sum^{k-3}_{j=1} {1\over j} \biggr]\ . \qquad\quad &(3.12b) \cr   }$$
Fortunately these coefficients are needed only for the particular set of
values $k=3$, $m=2$ and $\ell=0$, corresponding to the only term present
in (2.16) in this category, which is the antepenultimate term in (2.16c).
From the formulas (A.22) of Appendix A we find

$$_{3,2}\beta_0 = {1\over 6} \ , \qquad {}_{3,2}\gamma_0 = {7\over 72}
 \ .\eqno (3.13) $$

With the expression (3.8), the values of coefficients in Tables 1 and 2,
and the two values (3.13), we have the material for the computation of
the retarded integral of $T^{\alpha\beta}_{(M^2 M_{pq})}$.  The result
reads

$$\eqalignno{
{\rm FP} \square^{-1}_R \left[ M^{-2}
T^{00}_{(M^2 M_{pq})} \right] &= {n_{ab}\over r} \int^{+\infty}_0
d\tau M^{(5)}_{ab}
\left\{ -4\ln^2 \left( {\tau\over 2r} \right) - 24 \ln \left(
{\tau\over 2r} \right) - {154\over 3} \right\} \cr
&+ o \left( {r^{\varepsilon-2}} \right)  \ , &(3.14a) \cr\cr  
{\rm FP} \square^{-1}_R \left[ M^{-2} 
T^{0i}_{(M^2 M_{pq})} \right] &= {n_a\over r}
\int^{+\infty}_0 d\tau M^{(5)}_{ai}
\left\{ -4 \ln^2 \left( {\tau\over 2r} \right)  - 16 \ln
\left( {\tau\over 2r} \right) - {202\over 5} \right\} \cr
&+ {16\over 9} {\hat{n}_{iab}\over r}
M^{(4)}_{ab} + o \left( {r^{\varepsilon-2}} \right) \ , &(3.14b)  \cr\cr  
{\rm FP} \square^{-1}_R \left[ M^{-2}
T^{ij}_{(M^2 M_{pq})} \right] &= {1\over r}
\int^{+\infty}_0 d\tau M^{(5)}_{ij}
\left\{ -4 \ln^2 \left( {\tau\over 2r} \right) - {1\over 6}
\ln \left( {\tau\over 2r_0} \right) - {1217\over 120} \right\} \cr
&+{1\over r} \Bigl[ {23\over 30}
\hat{n}_{ijab} M^{(4)}_{ab} + {226\over 63} \delta_{ij} n_{ab}
M^{(4)}_{ab}
- {52\over 7} n_{a(i} M^{(4)}_{j)a} \Bigr] \cr
&+ o \left( {r^{\varepsilon-2}} \right) \ . &(3.14c)  \cr  }
$$
All the logarithms (and logarithms square) are computed with $\tau/2r$
except for one, in the last equation (3.14c), which is computed with
$\tau/2r_0$ and is issued from the formula (3.11).  The other
logarithms, and logarithms square, are issued from (3.8).
\section{The monopole-monopole-quadrupole metric}   
\subsection{The metric in the far zone} 

The retarded integral of the monopole-monopole-quadrupole source, namely

$$u^{\alpha\beta}_{(M^2 M_{pq})} = {\rm FP}
  \square^{-1}_R [\Lambda^{\alpha\beta}_{(M^2 M_{pq})}]\ , \eqno (4.1) $$
is obtained simply as the sum of (3.5) and (3.14). Let us now follow the
method proposed in [16] (see also Section 2 of paper I), and add to
$u^{\alpha\beta}_{(M^2 M_{pq})}$ a supplementary term
$v^{\alpha\beta}_{(M^2 M_{pq})}$ so designed as (i) to be a solution of
the homogeneous wave equation, and (ii) to be such that the sum of
$u^{\alpha\beta}_{(M^2 M_{pq})}$ and $v^{\alpha\beta}_{(M^2 M_{pq})}$ is
divergenceless.  With (i) and (ii) satisfied, a particular solution of
the cubic-order field equations in harmonic coordinates is
$u^{\alpha\beta}_{(M^2 M_{pq})} + v^{\alpha\beta}_{(M^2 M_{pq})}$.
Actually we do not follow exactly the construction proposed in [16], but
adopt the slighly modified construction of $v^{\alpha\beta}_{(M^2
M_{pq})}$ defined by the equations (2.11) and (2.12) of paper I.
Furthermore, as we control only the dominant behaviour at infinity of the
term $u^{\alpha\beta}_{(M^2 M_{pq})}$, we must check that this weaker
information still permits the construction from the divergence of
$u^{\alpha\beta}_{(M^2 M_{pq})}$ of the corresponding
$1/r$ term in $v^{\alpha\beta}_{(M^2 M_{pq})}$.  This poses no problem,
and the relevant formulas can be found in Appendix B.  The divergence of
$u^{\alpha\beta}_{(M^2 M_{pq})}$, computed from (3.5) and (3.14), reads

$$ \eqalignno{ 
\partial_\beta \left[ M^{-2} u^{0\beta}_{(M^2 M_{pq})}\right] &= {176\over
105} {n_{ab}\over r} M^{(5)}_{ab} + O \left( {1\over r^2} \right)\ , 
&(4.2a) \cr
\partial_\beta \left[ M^{-2} u^{i\beta}_{(M^2 M_{pq})}\right]  
 &= {n_a\over r}  \int^{+\infty}_0 d\tau
   M^{(6)}_{ia} \left\{ - {9\over 10}
\ln \left( {\tau\over 2r_0} \right) - {1211\over 600} \right\}\cr
 &+{36\over 35} {\hat{n}_{iab}\over r} M^{(5)}_{ab} + O
\left( {1\over r^2} \right) \ . &(4.2b) \cr }$$
Then the formulas (B.2)-(B.5) in Appendix B give the $1/r$ term in
$v^{\alpha\beta}_{(M^2 M_{pq})}$ as

$$  \eqalignno{
 M^{-2} v^{00}_{(M^2 M_{pq})} &= O \left( {1\over r^2} \right) \ , &(4.3a) \cr
 M^{-2} v^{0i}_{(M^2 M_{pq})} &= {176\over 105} {n_a\over r} 
    M^{(4)}_{ai} + O \left( {1\over r^2} \right)  \ , &(4.3b) \cr
 M^{-2} v^{ij}_{(M^2 M_{pq})} &= {1\over r} \int^{+\infty}_0 d\tau M^{(5)}_{ij}
  \left\{ -{9\over 10} \ln \left( {\tau\over 2r_0} \right)
  - {499\over 280} \right\} \cr
  &+{1\over r} \left[ -{72\over 35}
   \delta_{ij} n_{ab} M^{(4)}_{ab} + {216\over 35} \hat{n}_{a(i}
   M^{(4)}_{j)a} \right] + O \left( {1\over r^2} \right) \ .&(4.3c)  \cr }$$
The complete monopole-monopole-quadrupole metric, defined by

$$h^{\alpha\beta}_{(M^2 M_{pq})} = u^{\alpha\beta}_{(M^2 M_{pq})} +
   v^{\alpha\beta}_{(M^2 M_{pq})} \ ,  \eqno (4.4)  $$
is therefore obtained by adding up the expressions (3.5), (3.14) and
(4.3).  We obtain

$$\eqalignno{
M^{-2} h^{00}_{(M^2 M_{pq})} &= {n_{ab}\over r} \int^{+\infty}_0
  d\tau M^{(5)}_{ab} \biggl\{ -4 \ln^2 \left( {\tau\over 2r} \right) 
  -4 \ln \left( {\tau\over 2r} \right) \cr
  &\qquad\qquad + {116\over 21} \ln \left( {\tau\over 2r_0}
  \right) - {7136\over 2205} \biggr\}
  + o \left( {r^{\varepsilon-2}} \right) \ , &(4.5a)  \cr
M^{-2} h^{0i}_{(M^2 M_{pq})} &= {\hat{n}_{iab}\over r} \int^{+\infty}_0
  d\tau M^{(5)}_{ab} \biggl\{ -{2\over 3} \ln \left( {\tau\over 2r} \right)
  -{4\over 105} \ln \left( {\tau\over 2r_0} \right) - {716\over 1225}
  \biggr\} \cr
  &+ {n_a\over r} \int^{+\infty}_0 d\tau M^{(5)}_{ai}
  \biggl\{ -4 \ln^2 \left( {\tau\over 2r} \right) - {18\over 5}
  \ln \left( {\tau\over 2r} \right) \cr
  &\qquad\qquad + {416\over 75} \ln \left(
  {\tau\over 2r_0} \right) - {22724\over 7875} \biggr\} 
   + o \left( {r^{\varepsilon-2}} \right) \ , &(4.5b)  \cr
M^{-2} h^{ij}_{(M^2 M_{pq})} &= {\hat{n}_{ijab}\over r} \int^{+\infty}_0
  d\tau M^{(5)}_{ab} 
  \biggl\{ - \ln \left( {\tau\over 2r} \right)
 - {191\over 210}  \biggr\} \cr&+ {\delta_{ij} n_{ab}\over r}
  \int^{+\infty}_0 d\tau
 M^{(5)}_{ab} \biggl\{ -{80\over 21} \ln \left( {\tau\over 2r} \right) -
 {32\over 21} \ln \left( {\tau\over 2r_0} \right) - {296\over 35}
 \biggr\} \cr
 &+ {\hat{n}_{a(i}\over r} \int^{+\infty}_0 d\tau
 M^{(5)}_{j)a} \biggl\{ {52\over 7} \ln \left( {\tau\over 2r} \right) +
 {104\over 35} \ln \left( {\tau\over 2r_0} \right) + {8812\over 525}
 \biggr\} \cr
 &+ {1\over r} \int^{+\infty}_0 d\tau M^{(5)}_{ij}
 \biggl\{ -4 \ln^2 \left( {\tau\over 2r} \right) - {24\over 5}
 \ln \left( {\tau\over 2r} \right) \cr
 &\qquad\qquad + {76\over 15} \ln \left(
 {\tau\over 2r_0} \right) - {198\over 35}
 \biggr\} + o \left( {r^{\varepsilon-2}} \right) &(4.5c)  \cr }$$
(the moments in the integrands are evaluated at $t-r-\tau$).

\subsection{The observable quadrupole moment}   

The computation we have done so far uses harmonic coordinates, which are
convenient for constructing solutions by means of a post-Minkowskian
algorithm (essentially because all the components of the field obey some
wave equations).  However, the harmonic coordinates entail a small
disadvantage, namely the associated coordinate cones $t-r$ deviate by a
logarithm of $r$ (in first approximation) from the true light cones
along which gravity propagates.  As a result, the expansion of the
metric when $r \to \infty$, $t-r=$ const involves besides the normal
powers of $1/r$ some powers of the logarithm of $r$ [42,43,16,17].  This
is clear from the previous result (4.5).  The logarithms can be gauged
away by going to some radiative coordinates $X^\mu$, which are such
that the associated coordinate cones $T-R$ (where $R=|{\bf X}|$) become
asymptotically tangent to the true light cones at infinity.  In this
paper it will be sufficient to check that the coordinate transformation

$$ X^\mu = x^\mu + G \xi^\mu  \ , \eqno(4.6a) $$
where $x^\mu$ are the harmonic coordinates and $\xi^\mu$ is given by

$$ \xi^0 = - 2M \ln \left( {r\over r_0} \right) \ , \qquad
   \xi^i = 0  \ , \eqno(4.6b) $$
does remove all the logarithms of $r$ in the particular case of the
interaction $M^2 \times M_{pq}$ (and at leading order at infinity).
Note that we have introduced in the coordinate transformation (4.6) the
{\it same} constant $r_0$ as used in the definition of the finite part
process in Section 3.  Actually we could have introduced any constant
$r_1$.  However the choice $r_1=r_0$, which is simply a choice of gauge
(equivalent to a choice of the origin of time in the far zone), is
especially convenient, as it will simplify some formulas below.

Under the coordinate transformation (4.6) the metric is changed to

$$H^{\mu\nu}_{(M^2 M_{pq})}(X) = h^{\mu\nu}_{(M^2 M_{pq})} (X) -
\xi^\lambda \partial_\lambda h^{\mu\nu}_{(M M_{pq})} + {1\over 2}
\xi^\lambda \xi^\sigma \partial_{\lambda\sigma} h^{\mu\nu}_{(M_{pq})}
  + o \left( {R^{\varepsilon-2}} \right) \ , \eqno (4.7)
$$
where we keep only the terms corresponding to the interaction $M^2
\times M_{pq}$ and neglect all sub-dominant terms
$o(R^{\varepsilon-2})$.  Both sides of (4.7) are expressed with the
radiative coordinates $X^\mu$.  We substitute in the right side the
harmonic-coordinates linear metric (2.5), quadratic one (2.8), and cubic
(4.5), and find that all logarithms disappear to order $1/R$, so that we
obtain the ``radiative'' metric

$$\eqalignno{
M^{-2} H^{00}_{(M^2 M_{pq})} &= {N_{ab}\over R} \int^{+\infty}_0 d
  \tau M^{(5)}_{ab} \left( T_R - \tau \right)
  \left\{ - 4 \ln^2 \left( {\tau\over 2r_0} \right) + {32\over 21}
  \ln \left( {\tau\over 2r_0} \right) - {7136\over 2205} \right\} \cr
  &+ o \left( {R^{\varepsilon-2}} \right) \ , &(4.8a) \cr
M^{-2} H^{0i}_{(M^2 M_{pq})} &= {\hat{N}_{iab}\over R} \int^{+\infty}_0 d
  \tau M^{(5)}_{ab} \left( T_R - \tau \right)
  \left\{ - {74\over 105} \ln \left( {\tau\over 2r_0} \right) -
  {716\over 1225} \right\} \cr
  &+ {N_a\over R} \int^{+\infty}_0 d
  \tau M^{(5)}_{ai} \left( T_R - \tau \right)
  \left\{ - 4 \ln^2 \left( {\tau\over 2r_0} \right) + {146\over 75}
  \ln \left( {\tau\over 2r_0} \right) - {22724\over 7875} \right\} \cr &+ 
  o \left( {R^{\varepsilon-2}} \right) \ , &(4.8b) \cr
M^{-2} H^{ij}_{(M^2 M_{pq})} &= {\hat{N}_{ijab}\over R} \int^{+\infty}_0 d
  \tau M^{(5)}_{ab} \left( T_R - \tau \right)
   \left\{ - \ln \left( {\tau\over 2r_0} \right) -
  {191\over 210} \right\} \cr
  &+ {\delta_{ij} N_{ab}\over R} \int^{+\infty}_0 d
  \tau M^{(5)}_{ab} \left( T_R - \tau \right)
   \left\{ - {16\over 3} \ln \left( {\tau\over 2r_0} \right)
  - {296\over 35}  \right\} \cr
  &+ {\hat{N}_{a(i}\over R} \int^{+\infty}_0 d
   \tau M^{(5)}_{j)a} \left( T_R - \tau \right)
   \left\{  {52\over 5} \ln \left( {\tau\over 2r_0} \right) +
  {8812\over 525} \right\} \cr
  &+ {1\over R} \int^{+\infty}_0 d
  \tau M^{(5)}_{ij} \left( T_R - \tau \right)
  \left\{ - 4 \ln^2 \left( {\tau\over 2r_0} \right) + {4\over 15}
  \ln \left( {\tau\over 2r_0} \right) - {198\over 35} \right\} \cr
  &+ o \left( {R^{\varepsilon-2}} \right) \ . &(4.8c) \cr  }$$
The retarded time is denoted by $T_R=T-R$ and the direction to the
observer by $N_a=N^a=X^a/R$.  Note that

$$K_\nu H^{\mu\nu}_{(M^2 M_{pq})} = o \left( {R^{\varepsilon-2}} \right) \ ,
      \eqno (4.9) $$
where $K_\nu=(-1,N^i)$ is the Minkowskian null direction to the observer.

From the radiative metric (4.8) we extract the ``observable'' multipole
moments which are the quantities measured by an observer located at infinity.
The observable multipole moments $U_L$ and $V_L$ parametrize the
algebraic transverse-tracefree (TT) projection of the spatial metric in
radiative coordinates,

$$ \eqalignno{
 (H^{ij})_{\rm TT} =& - {4\over R} {\cal P}_{ijab}
 \sum^\infty_{\ell =2} {1\over \ell !} \biggl\{ N_{L-2} U_{ijL-2}(T_R)
 - {2\ell \over {\ell +1}} N_{aL-2}
 \varepsilon_{ab(i} V_{j)bL-2}(T_R) \biggr\} \cr
  &+ O \left( {1\over R^2}\right) \ , &(4.10) \cr } $$
where the TT projection operator is

$$ {\cal P}_{ijab} ({\bf N}) = (\delta_{ia} -N_iN_a) (\delta_{jb} -N_jN_b)
    -{1\over 2} (\delta_{ij} -N_iN_j) (\delta_{ab} -N_aN_b) \ .\eqno(4.11) $$
[As we use the field variable $H^{ij}$ instead of the covariant metric,
the formula (4.10) differs by a sign from (5.1) in paper I.] The moments
$U_L$ and $V_L$ agree at the linearized order with the $\ell$th
time-derivatives of the moments $M_L$ and $S_ L$ [14].

Working out the TT projection and comparing the result with (4.10), we
readily find that the monopole-monopole-quadrupole metric (4.8)
contributes to the {\it quadrupole} observable moment $U_{ij}$, and only
to this moment, by the expression

$$ \delta U_{ij} (T_R) = 2 M^2 \int^{+\infty}_0
   d \tau M^{(5)}_{ij} \left( T_R - \tau \right)
  \left\{ \ln^2 \left( {\tau\over 2r_0} \right) + {57\over 70} \ln
  \left( {\tau\over 2r_0} \right) + {124627\over 44100} \right\} \ .
    \eqno (4.12) $$
We add back the necessary powers of $G$ and $1/c$ [recall that $\delta
U_{ij}$ has the dimension of (mass)(length/time)$^2$], and find that the
term (4.12) carries in front a factor $G^2/c^6$, and therefore
represents a small modification of the lowest-order quadrupole
radiation at the level of the third post-Newtonian (3PN) order.  Let us
prove that there is no other contributions in $U_{ij}$ at this level.
We know from dimensional arguments that a non-linear term in $U_{ij}$
involves necessarily a factor $1/c^{3(n-1)+ \Sigma\ell_i+s-2}$, where
$n \geq 2 $ is the order of non-linearity, where $\Sigma\ell_i$ is the
sum of multipolarities of the $n$ moments $M_L$ and/or $S_L$ composing
the term ($1 \leq i \leq n$), and where $s$ is the number of current-type
moments $S_L$ (see for instance Section V in [35]).
Furthermore, $\Sigma\ell_i+s-2=2k$ where $k$ is the number of
contractions of indices among all the indices carried by the moments.
For a term at the 3PN order we thus have $3(n-1)+2k=6$, which has the
unique solution $n=3$ (cubic interaction) and $k=0$ (no contraction of
indices). Then the multipolarities satisfy $\ell_1+\ell_2+\ell_3+s-2=0$,
so we have necessarily $(\ell_1,\ell_2,\ell_3)=(0,0,2)$ and $s=0$, which
corresponds indeed to the sole interaction $M^2 \times M_{ij}$.

The equation (5.10) of paper I gives the observable quadrupole moment
$U_{ij}$ including all terms up to the 2.5PN order.  Therefore, by the
previous reasoning, we can simply add to (5.10) of paper I the
contribution of tails of tails at 3PN order and obtain the complete
$U_{ij}$ to the 3PN order.  We have

$$\eqalignno{
U_{ij}(T_R) &= M^{(2)}_{ij} + 2 {GM\over c^3} \int^{+\infty}_0 d
\tau M^{(4)}_{ij} (T_R-\tau) \left[ \ln \left( {c\tau\over 2r_0} \right)
+ {11\over 12} \right] \cr
&+ {G\over c^5} \Biggl\{ - {2\over 7} \int^{+\infty}_0 d \tau
\left[M^{(3)}_{a<i} M^{(3)}_{j>a}\right](T_R-\tau) - {2\over 7} M^{(3)}_{a<i}
M^{(2)}_{j>a}  \cr
&- {5\over 7} M^{(4)}_{a<i}  M^{(1)}_{j>a}  + {1\over 7}
M^{(5)}_{a<i} M_{j>a} + {1 \over 3} \varepsilon_{ab<i}
M^{(4)}_{j>a} S_b \Biggr\} \cr
&+ 2 \left( {GM\over c^3} \right)^2 \int^{+\infty}_0 d
\tau M^{(5)}_{ij}(T_R-\tau)
\left[ \ln^2 \left( {c\tau\over 2r_0} \right) + {57\over 70}
\ln \left( {c\tau\over 2r_0} \right) + {124627\over 44100} \right] \cr
&+ O \left( {1\over c^7} \right) \ . &(4.13) \cr  }$$
The various terms are:  at 1.5PN order, the dominant tail integral [32];
at 2.5PN order, the quadrupole-quadrupole terms [including in particular
a non-local (memory) integral] and a quadrupole-dipole term (see paper I
and references therein);  and, at 3PN order, the tail of tail integral
computed in this paper. The formula (4.13) constitutes the main result of
this paper, as it gives all the physical effects in the radiation field
measured by a far away detector up to the 3PN order.

Note that $U_{ij}$, when expressed in terms of the intermediate moments
$M_L$ and $S_L$ as in (4.13), shows a dependence on the (arbitrary)
length scale $r_0$.  Most of this dependence comes from our definition
(4.6) of a radiative coordinate system, and thus can be removed by
inserting $T_R = t-r/c-(2GM/c^3){\rm ln}(r/r_0)$ back into (4.13), and
expanding the result when $c \to \infty$, keeping the necessary terms
consistently.  In doing so one finds that there remains a
$r_0$-dependent term at 3PN order, namely

$$ U_{ij}=M^{(2)}_{ij}- {214\over 105} \ln\left(r \over r_0\right)
   \left( {GM\over c^3} \right)^2
   M^{(4)}_{ij}+\hbox{terms independent of} ~r_0 \ .  \eqno (4.14a) $$
This term results simply from our use of the $r_0$-dependent formulas (3.4)
and (3.11) in constructing the harmonic-coordinates metric.  As we see
from (4.14a), the dependence of $U_{ij}$ on $r_0$ (or rather $r_0/c$) is
through the effective quadrupole moment

$$ M^{\rm eff}_{ij} = M_{ij}+{214\over 105} \ln\left(r_0\over c\right)
    \left( {GM\over c^3} \right)^2 M^{(2)}_{ij} \ .  \eqno (4.14b) $$
This moment is exactly the one which appears in the near-zone expansion
of the external metric, when taking into account the appearance of the
dominant logarithm of $c$ arising at the 3PN approximation. See the
discussion in the Appendix of [29]. The appearance of this $\ln c$ was pointed out by Anderson {\it et al} [44].

\subsection{The energy flux} 

We now investigate the total energy flux generated by the isolated source
(its total gravitational {\it luminosity}), and notably the non-local
contributions therein.  From (4.10) the
luminosity in terms of the observable moments reads

$$
 {\cal L} = \sum^{+\infty}_{\ell=2} {G\over c^{2\ell+1}} \left\{
   {(\ell+1)(\ell+2)\over (\ell-1)\ell \ell!(2\ell+1)!!} U^{(1)}_L
   U^{(1)}_L + {4\ell (\ell+2)\over (\ell-1)(\ell+1)!(2\ell+1)!!c^2}
   V^{(1)}_L V^{(1)}_L \right\}  \ .\eqno(4.15)
$$
The powers of $1/c$ show notably that the moments $U_L$ and $V_L$ of
higher multipolarities $\ell$ contribute to higher orders in the
post-Newtonian expansion of ${\cal L}$.

Let us gather the available information on non-local effects present in
the $U_L$'s and $V_L$'s.  Rather, we consider the time-derivatives of
the moments ($U^{(1)}_L$ and $V^{(1)}_L$), because these are the
quantities of interest in (4.15).  From (4.13) we write the
time-derivative of the quadrupole $U_{ij}$ as

$$\eqalignno{
 U^{(1)}_{ij}&=M^{(3)}_{ij}+2{GM\over c^3} \int^{+\infty}_0 d \tau
M^{(5)}_{ij} \left[\ln \left({c\tau\over 2r_0}\right)
+{11\over 12}\right] \cr
&+{1\over c^5} \biggl\{\hbox{instantaneous terms}~ M_{a<i}
M_{j>a}, \hbox{and}~ \varepsilon_{ab<i} M_{j>a} S_b \biggr\} \cr
&+2\left({GM\over c^3}\right)^2
\int^{+\infty}_0 d\tau M^{(6)}_{ij} \left[\ln^2 \left({c\tau\over
2r_0}\right) +{57\over 70} \ln \left({c\tau \over 2r_0}\right)
+{124627\over 44100}\right] \cr
&+{1\over c^7} \biggl\{\hbox{instantaneous terms}~ M_{ab<i}
M_{j>ab}, M_{ab} M_{ijab},
\hbox{and} ~\varepsilon_{ab<i} M_{j>ac} S_{bc} \biggr\}\cr
&+O\left({1\over c^8}\right) \ , &(4.16) \cr}
$$
where we indicate only the index {\it structure} of the local
(instantaneous) terms, but write {\it in extenso} all the non-local
integrals.  Note the important fact that the non-local (memory) integral
present in $U_{ij}$ at the 2.5PN order is a mere time anti-derivative
[see (4.13)], and therefore becomes instantaneous when considering the
time-derivative.  We have added with respect to (4.13) the information
that the 3.5PN term is instantaneous, exactly like the 2.5PN term.  This
follows from the dimensional argument used before.  The 3.5PN term is
such that $3(n-1)+2k=7$, therefore $n=2$ and $k=2$.  Now we know [32]
that the only non-local integrals at the quadratic order $n=2$ are the
tail integral which is purely of order 1.5PN, and the memory integral
which contributes to the 2.5PN, 3.5PN and higher orders but is in the
form of a simple time anti-derivative of an instantaneous functional of
the moments $M_L$ and $S_L$. This proves that the 3.5PN term in (4.16) is
indeed instantaneous. Furthermore its multipole structure follows from
$\ell_1+\ell_2+s-2=2k=4$. Note that the instantaneous terms in (4.16)
(and other equations below) are instantaneous functionals not only of
the moments $M_L$ and $S_L$ but also of the real source variables, i.e.
when $M_L$ and $S_L$ are replaced by their explicit expressions as integrals
over the source.  Indeed the non-local integrals in $M_L$ and $S_L$ are not
expected to arise before the 4PN order.

Similarly we write the relevant higher-order multipole moments, but for
them we need less accuracy than for the quadrupole.  The results are

$$\eqalignno{
  U^{(1)}_{ijk}&=M^{(4)}_{ijk}+2{GM\over c^3} \int^{+\infty}_0 d\tau
  M^{(6)}_{ijk} \left[\ln \left({c\tau\over 2r_0}\right)
  +{97\over 60}\right] \cr
&+{1\over c^5} \biggl\{ \hbox{instantaneous terms}~ M_{a<ij} M_{k>a}\ ,
 \ \varepsilon_{ab<i} M_{jk>a} S_b\ ,\cr
&\qquad\quad \hbox{and}\ \varepsilon_{ab<i}
M_{j {\underline{a}}} S_{k>b} \biggr\}+O\left({1\over c^6}\right)\ ,&(4.17a)\cr
 V^{(1)}_{ij}&=S^{(3)}_{ij}+2{GM\over c^3} \int^{+\infty}_0 d\tau\
 S^{(5)}_{ij}
 \left[\ln \left({c\tau\over 2r_0}\right)+{7\over 6} \right]\cr
&+{1\over c^5} \biggl\{ \hbox{instantaneous terms}\ M_{a<i} S_{j>a},\
M_{aij} S_a,\ \varepsilon_{ab<i} M_{j>ac} M_{bc} \ , \cr
&\qquad\quad \hbox{and}\ \varepsilon_{ab<i} S_{j>a} S_b \biggr\}
+O\left({1\over c^6}\right)\ ,  &(4.17b) \cr
U^{(1)}_{ijkl}&=M^{(5)}_{ijkl}+{G\over c^3} \biggl\{2M \int^{+\infty}_0
d\tau M^{(7)}_{ijkl} \left[\ln \left({c\tau \over 2r_0}\right)
+{59\over 30}\right] \cr
&\qquad +\hbox{instantaneous terms}\ M_{<ij} M_{kl>} \biggr\}
+O \left({1\over c^4}\right)\ , 
&(4.17c) \cr 
V^{(1)}_{ijk}&=S^{(4)}_{ijk}+{G\over c^3} \biggl\{ 2M \int^{+\infty}_0
d \tau~ S^{(6)}_{ijk} \left[\ln \left({c\tau \over 2r_0}\right)
+{5\over 3}\right]  \cr
&\qquad + \hbox{instantaneous terms}\ M_{<ij} S_{k>},\ \hbox{and}\
\varepsilon_{ab<i}
M_{j {\underline{a}}} M_{k>b} \biggr\} +O \left({1\over c^4}\right) \ .\cr
& &(4.17d) \cr}
$$
[The tails in the mass octupole and current quadrupole are from (5.8) in
[35].  The mass $2^4$-pole and current octupole are computed (including
all the instantaneous terms) in (5.11) and (5.12) of paper I.]

The separation made in the moments (4.16)-(4.17) between instantaneous and
non-local terms yields a similar separation in the energy flux given by (4.15).
Furthermore we introduce in the non-local part of ${\cal L}$ a separation
between the tail terms strictly speaking, a term involving the square of
the tail, and the tail of tail term. Accordingly, we denote

$$ {\cal L}={\cal L}_{\rm inst}+{\cal L}_{\rm tail}+{\cal L}_{(\rm tail)^2}
    +{\cal L}_{\rm tail (\rm tail)} \ . \eqno(4.18) $$
Quite evidently we include into the tail part of the flux, ${\cal
L}_{\rm tail}$, all the terms which are made of the cross products of
the moments $M_L$ and $S_L$ and of the corresponding tail integrals at
1.5PN order.  From (4.15)-(4.17) we obtain, neglecting terms at the 4PN
order,

$$\eqalignno{
{\cal L}_{\rm tail}={4G^2M\over c^5} &\left\{ {1\over 5c^3} M^{(3)}_{ij}
\int^{+\infty}_0 d\tau M^{(5)}_{ij} (T_R-\tau) \left[\ln
\left({c\tau \over 2r_0}\right)
+{11\over 12}\right] \right.\cr
&+{1\over 189c^5}~ M^{(4)}_{ijk} \int^{+\infty}_0 d\tau M^{(6)}_{ijk}
(T_R-\tau) \left[\ln \left({c\tau\over 2r_0}\right)+{97\over 60} \right] \cr
&+{16\over 45c^5}~S^{(3)}_{ij} \int^{+\infty}_0 d\tau~ S^{(5)}_{ij}
(T_R-\tau) \left[\ln \left({c\tau\over 2r_0}\right)+{7\over 6}\right]  \cr
&+{1\over 9072c^7}~ M^{(5)}_{ijkl} \int^{+\infty}_0 d\tau~ M^{(7)}_{ijkl}
(T_R-\tau) \left[\ln \left({c\tau\over 2r_0}\right)+{59\over 30}\right] \cr
&+{1\over 84c^7}~ S^{(4)}_{ijk} \int^{+\infty}_0 d\tau~ S^{(6)}_{ijk}
(T_R-\tau) \left[\ln \left({c\tau\over 2r_0}\right)+{5\over 3}\right]
\cr
&\left.+O\left({1\over c^8}\right) \right\}  \ . &(4.19)\cr}
$$
The moments in front of each integrals depend on the current time $T_R$.
[For convenience we include in (4.19) the terms associated with the
constants $11/12$, $97/60$, etc., though these terms are actually
instantaneous.  In fact these terms are given by some total
time-derivatives and thus do not participate to the loss of energy in the
source.  In the case of binary systems moving on circular orbits,
these terms are rigorously zero (see Section 5).] The expression (4.19)
generalizes to 3.5PN order the expression (5.12) of [35].  

Now the ``(tail)$^2$'' contribution to the flux is given by the square of
the tail integral at 1.5PN order, and therefore enters the energy flux at
the same 3PN order as the contribution of tails of tails. In fact this
contribution could be treated on the same footing as the ``tail(tail)''
contribution, but it will be clearer in Section 5 to investigate it
separately.  We have

$$\eqalignno{
 {\cal L}_{(\rm tail)^2}={4G^2M\over c^5} &\left\{ {GM\over 5c^6}
\left(\int^{+\infty}_0 d\tau M^{(5)}_{ij} (T_R-\tau)
\left[\ln \left({c\tau\over 2r_0}\right)+{11\over 12}\right] \right)^2 \right.\cr
&\left.+O\left({1\over c^{10}}\right)\right\} \ . &(4.20)\cr}
$$
Physically (4.20) represents the energy flux due to the tail part of the
wave, in situations where the tail can be separated from the other
components of the field.  In particular, after the passage of a burst of
gravitational radiation (defined by the constancy of the quadrupole
moment $M_{ij}$ before and after a certain interval of time), the wave
tail will be solely present in the radiation field, and therefore the
total energy flux ${\cal L}$ will reduce in this case to ${\cal L}_{(\rm
tail)^2}$.

Next, the ``tail(tail)'' contribution to the flux involves the cross
product of the quadrupole moment and of the tail of tail integral at 3PN
order.  It reads (neglecting 4PN-order terms)

$$\eqalignno{
 {\cal L}_{\rm tail (\rm tail)} ={4G^2M\over c^5} &\left\{ {GM\over 5c^6}
 M^{(3)}_{ij} \int^{+\infty}_0 d\tau M^{(6)}_{ij} (T_R-\tau) \right.\cr
&\left.\times \left[\ln^2 \left({c\tau\over 2r_0}\right) + {57\over 70} \ln
 \left({c\tau \over 2r_0}\right) + {124627\over 44100}\right]
+O\left({1\over c^8}\right)\right\} \ . \cr &  &(4.21)}
$$
Note that in contrast to (4.19) the instantaneous terms associated with the
constants $11/12$ and $124627/44100$ in (4.20) and (4.21) do contribute to
the energy flux, even in the case of binary systems moving on circular
orbits (see Section 5).

Finally the instantaneous part of the flux, ${\cal L}_{\rm inst}$, is
entirely defined, up to the 3.5PN order included, by the general formula
(4.15) together with the previous definitions of ${\cal L}_{\rm tail}$,
${\cal L}_{(\rm tail)^2}$ and ${\cal L}_{\rm tail(\rm tail)}$. We do
not write the full expression of ${\cal L}_{\rm inst}$ in terms of $M_L$
and $S_L$ because we do not consider it in the application to compact
binaries in Section 5. It suffices for our purpose to recall that
${\cal L}_{\rm inst}$ not only is instantaneous in terms of the moments
$M_L$ and $S_L$, but also is instantaneous in terms of the real source
parameters (neglecting 4PN-order terms), that is, in the case of compact
binaries, of the orbital separation and relative velocity of the two
bodies.

\section{Application to inspiraling compact binaries}  

Last but not least we specialize the results to binary systems of compact
objects (neutron stars or black holes). These systems, when the two objects
spiral very rapidly toward each other in the phase just prior to the
final coalescence (the orbital motion is highly relativistic in this
phase), constitute the most interesting known source of gravitational
waves to be observed by VIRGO and LIGO.  The rate of inspiral is fixed
by the total energy in the gravitational waves generated by the orbital
motion, that is by the binary's total gravitational luminosity $\cal L$,
which is therefore a crucial quantity to predict.  We assume two
non-spinning point masses (without internal structure) moving on an
orbit which evolved for a sufficiently long time to have been
circularized by the radiation reaction forces.  For such (excellent)
modelling of inspiraling compact binaries we compute ${\cal L}_{\rm
tail}$, ${\cal L}_{(\rm tail)^2}$ and ${\cal L}_{\rm tail(\rm tail)}$ to
the 3.5PN order included.

To compute the tail part [given by (4.19)] we need the expressions of
the multipole moments for circular compact binaries, to 2PN order for
the mass quadrupole moment $M_{ij}$, 1PN order for the mass octupole
$M_{ijk}$ and current quadrupole $S_{ij}$, and Newtonian order for
$M_{ijkl}$ and $S_{ijk}$ (we need also the ADM mass $M$ to 2PN
order). These moments have been calculated in [45], and we simply
report here their expressions:

$$\eqalignno{ M&=m \left[1-{\gamma\over 2} \nu + {\gamma^2\over
8}(7\nu -\nu^2)+O\left({1\over c^5}\right)\right]\ , &(5.1a)\cr M_{ij}&=\nu
m \biggl\{x^{<ij>} \left[1-{\gamma\over 42} (1+39\nu)- {\gamma^2\over
1512}(461+18395\nu +241\nu^2)\right]\cr &\qquad +{r^2\over c^2}
v^{<ij>}
\left[ {11\over 21}(1-3\nu) +{\gamma\over 378} (1607-1681\nu +229\nu^2)\right]
+O\left({1\over c^5}\right)\biggr\} \ , \cr 
& &(5.1b)\cr
M_{ijk}&=-\nu \delta m \left\{x^{<ijk>} (1-\gamma \nu)+{r^2\over c^2}
v^{<ij}x^{k>} (1-2\nu)+O\left({1\over c^4}\right) \right\}\ ,  &(5.1c)\cr
S_{ij}&= -\nu \delta m~ \varepsilon^{ab<i} x^{j>a} v^b
\left[1+{\gamma\over 28}
(67-8\nu)+O\left({1\over c^4}\right)\right] \ , &(5.1d) \cr
M_{ijkl}&=\nu m x^{<ijkl>} (1-3\nu)+O\left({1\over c^2}\right)\ , &(5.1e) \cr
S_{ijk}&=\nu m~ \varepsilon^{ab<i} x^{jk>a} v^b (1-3\nu)
+O\left({1\over c^2}\right)\ .  &(5.1f)\cr}
$$
The mass parameters are the total mass $m=m_1+m_2$, the mass difference
$\delta m=m_1-m_2$, and the mass ratio $\nu=m_1m_2/m^2$ (satisfying $0 <
\nu \leq 1/4$).  The relative position and velocity of the two
point-masses are denoted by $x^i=y^i_1-y^i _2$ and $v^i=dx^i/dt$.  We
use the post-Newtonian parameter

$$ \gamma = {Gm \over {rc^2}} \ , \qquad r=|{\bf x}| \eqno(5.2) $$
($r$ is the harmonic-coordinates distance between the two bodies).  The
time-derivatives of the multipole moments (5.1) are computed using the
equations of motion, with maximal 2PN-precision needed for the
time-derivatives of the quadrupole moment.  The equations of motion are

$$\eqalignno{
{dv^i \over dt} &= -\omega^ 2_{2{\rm PN}} x^i + O \left( {1\over c^5} \right)
 \ , &(5.3a)\cr
\omega^2_{2{\rm PN}} &={Gm\over r^3} \left[1+(-3+\nu)\gamma
+\left(6+{41\over 4}\nu +\nu^2 \right)\gamma^2\right] \ , &(5.3b)\cr}
$$
where the frequency $\omega_{2{\rm PN}}$ is the orbital frequency of the
exact circular periodic motion at the 2PN order (see [45]). Introducing
instead of $\gamma$ the post-Newtonian parameter
$x=(Gm \omega_{2{\rm PN}}/c^3)^{2/3}$ we have

$$ \gamma = x \left\{1+\left(1-{\nu \over 3}\right)x 
+\left(1-{{65\nu}\over 12}\right)x^2+O\left(x^3\right)\right\}\ .\eqno(5.3c) $$
The computation of ${\cal L}_{\rm tail}$ proceeds like in Section VI.B of
[35].  Namely we insert into (4.19) the time-derivatives of the moments
(5.1), evaluated both at the current time $T_R$ and at all anterior
times $T_R-\tau$.  Then we work out all contractions of indices and
obtain many integrals of logarithms of $c\tau / 2 r_1$, where $r_1$ is
some constant such as $r_1=r_0 e^{-11/12}$, multiplied by cosines of
some multiples ($n$) of $\omega_{2{\rm PN}} \tau$.  All these integrals
are computed using the mathematical formula

$$
\int^{+\infty}_0 dy\ln y\, e^{-\lambda y}= -{1\over \lambda}(C+\ln \lambda)
\ , \eqno(5.4)
$$
where $\lambda$ denotes the complex number $\lambda=2$i$n\omega_{2{\rm
PN}} r_1/c$, and where $C=0.577..$ is the Euler constant (see e.g.  [47]
p.  573). See Appendix A of [30] for the proof that this formula yields
the correct result for inspiraling compact binaries modulo some error
terms of order $O(c^{-5}\ln c)$, falling in the present case into the
un-controlled remainder of (4.19).  In fact we need only the real part
of (5.4) for the computation of ${\cal L}_{\rm tail}$, which leads
[using $\ln ({\rm i}\omega)=\ln \omega+{\rm i}\pi/2$] to a term
proportional to $\pi$ and independent of $r_1$.  The result (extending
the equation (6.18) in [33]) is

$$\eqalignno{
  {\cal L}_{\rm tail}={32c^5\over 5G} \nu^2 \gamma^5 &\left\{4\pi
  \gamma^{3/2}-\left({25663\over 672}+{125\over 8}\nu\right)
  \pi \gamma^{5/2} \right.\cr
 &\left.+\left({90205\over 576}+{505747\over 1512}\nu
  +{12809\over 756}\nu^2\right)\pi\gamma^{7/2}
  +O(\gamma^4)\right\} \ ,  &(5.5a) \cr}
$$
or, equivalently, in terms of the parameter $x$,

$$\eqalignno{
  {\cal L}_{\rm tail}={32c^5\over 5G} \nu^2 x^5 &\left\{4\pi
  x^{3/2}-\left({8191\over 672}+{583\over 24}\nu\right)
  \pi x^{5/2} \right.\cr
 &\left.+\left(-{16285\over 504}+{214745\over 1728}\nu
  +{193385\over 3024}\nu^2\right)\pi x^{7/2}
  +O(x^4)\right\} \ .  &(5.5b) \cr}
$$
The tails contribute only to the {\it half-integer} post-Newtonian
approximations 1.5PN, 2.5PN, and 3.5PN. Now, we know that only the terms
given by some {\it non-local} integrals can contribute to the
half-integer post-Newtonian approximations.  This follows from an
argument presented in Sect.  VI B of [33], which shows that the terms
given by instantaneous functionals of the binary's relative position and
velocity are zero for half-integer approximations in the energy flux for
circular orbits (but only in this case).  Thus we conclude that the
terms computed in (5.5a) and (5.5b) represent the {\it complete} 1.5PN,
2.5PN, and 3.5PN approximations in $\cal L$ -- no other contributions
can come from ${\cal L}_{\rm inst}$ to these orders (and ${\cal L}_{(\rm
tail)^2}$ and ${\cal L}_{\rm tail (tail)}$ are purely of 3PN order). Being
complete these approximations can thus be compared, in the test-mass
limit $\nu \to 0$ for one body, with the result of black-hole perturbation
theory derived up to 4PN order by Tagoshi and Sasaki [50].  For the
comparison we must use (5.5b) expressed in a coordinate-independent way
by means of the parameter $x$.  We find that there is perfect agreement,
in the limit $\nu \to 0$, between (5.5b) and the corresponding terms in
the equation (43) of [50] (see also the equation (3.1) of [51]).

Turn now to ${\cal L}_{(\rm tail)^2}$ defined by (4.20). The computation
is essentially the same as for ${\cal L}_{\rm tail}$, but since we are
considering a small 3PN effect the expression of the quadrupole moment
at the Newtonian order is sufficient.  Consistently we use the Newtonian
equations of motion (with orbital frequency $\omega^2=Gm/r^3$).  Using
the mathematical formula (5.4), in which both the real and imaginary
parts are now needed, we readily obtain

$$\eqalignno{
  {\cal L}_{(\rm tail)^2}={32c^5\over 5G} \nu^2 \gamma^5 &\biggl\{
\left(16\left[C+\ln
  \left({4\omega r_0} \over c\right)\right]^2
 - {88\over 3} \left[C+\ln
  \left({4\omega r_0} \over c\right)\right]\right.\cr
&\left. \qquad+4\pi^2 
  +{121\over 9}\right)\gamma^3
+O(\gamma^4)\biggr\} \ .  &(5.6) \cr}
$$
Finally we compute ${\cal L}_{\rm tail(\rm tail)}$ as given by (4.21).
Again we need only the Newtonian quadrupole moment and Newtonian equations
of motion, however a new ingredient is necessary, which is a
mathematical formula analogous to (5.4) but able to deal with the {\it
square} of the logarithm. From [47] p.  574 the relevant formula is

$$ \int^{+\infty}_0 dy \ln^2y~e^{-\lambda y}={1\over \lambda}
   \left[{\pi^2\over 6} +(C+\ln \lambda)^2\right] \ . \eqno(5.7) $$
Using this formula for inspiraling compact binaries is justified in the
same way as for the earlier formula (5.4). As a result we get 

$$\eqalignno{
  {\cal L}_{\rm tail(\rm tail)}={32c^5\over 5G} \nu^2 \gamma^5 &\biggl\{
\left(-16\left[C+\ln
  \left({4\omega r_0} \over c\right)\right]^2
 + {456\over 35} \left[C+\ln
  \left({4\omega r_0} \over c\right)\right]\right.\cr
 &\left.\qquad+{4\over 3}\pi^2 
  -{498508\over 11025}\right)\gamma^3
+O(\gamma^4)\biggr\} \ .  &(5.8) \cr}
$$

Both ${\cal L}_{(\rm tail)^2}$ and ${\cal L}_{\rm tail(\rm tail)}$ have
the same structure, namely that of a polynomial of the second degree in
the combination $C+\ln (4\omega r_0/c)$.  However, we see that the
coefficients in front of the square of $C+\ln (4\omega r_0/c)$ in (5.6)
and (5.8) are exactly opposite, and therefore that the sum of ${\cal
L}_{(\rm tail)^2}$ and ${\cal L}_{\rm tail(\rm tail)}$ is actually
merely linear in the combination $C+\ln (4\omega r_0/c)$.  This fact is
somewhat surprising because the terms involving the square of
$\ln\omega$ are {\it a priori} allowed. Thus, to the 3PN order, the terms
$(\ln\omega)^2$ cancel out and there remains simply a term with $\ln\omega$
(and the associated $\ln c$).  The appearance of a $\ln\omega$ at 3PN
was first shown in this context by Tagoshi and Nakamura [39].  Recall
that the general structure of the post-Newtonian expansion of the
near-zone metric involves besides the regular powers of $1/c$ some
arbitrary powers of $\ln c$ [16]. Similarly one expects that the
general structure of ${\cal L}$ should involve when going to higher
post-Newtonian approximations some arbitrary powers of $\ln\omega$.
[Note that up to the 5.5PN order ${\cal L}$ as computed in the limit
$\nu \to 0$ by Tanaka {\it et al} [50] is still linear in $\ln\omega$.]

Adding up (5.6) and (5.8) we obtain

$$\eqalignno{
  {\cal L}_{(\rm tail)^2+\rm tail(\rm tail)}={32c^5\over 5G} \nu^2 \gamma^5
 &\biggl\{ \left( - {1712\over 105} \left[C+\ln
  \left({4\omega r_0} \over c\right)\right]
 +{16\over 3}\pi^2 -{116761\over 3675}\right)\gamma^3\cr
&\qquad+O(\gamma^4)\biggr\} \ .  &(5.9) \cr}
$$
It is not yet possible to compare this result when $\nu \to 0$ with the
one obtained using the perturbation theory. Indeed (5.9) represents only
a part of the complete 3PN term in $\cal L$, which should also take into
account the (hard to get) 3PN contributions in ${\cal L}_{\rm inst}$,
which are due to the 3PN relativistic corrections in the
quadrupole moment (5.1a) and in the equations of motion (5.3).  (The 3PN
contributions in ${\cal L}_{\rm inst}$ are in progress [56,57].)
Nevertheless, we already recognize in (5.9) the same terms with $\pi^2$
and the combination $C+\ln (4\omega)$ as in the result of Tagoshi and
Sasaki [50].  [To the 3PN order one can replace in (5.9) the parameter
$\gamma$ by $x$.]

The 3PN term obtained in (5.9) depends on the arbitrary length scale
$r_0$.  This is not a problem because the 3PN term in ${\cal L}_{\rm
inst}$ is expected also to depend on $r_0$ through in particular the
explicit expression of the intermediate quadrupole moment $M_{ij}$ as a
functional of the source's stress-energy tensor.  The $r_0$-dependent
terms in both ${\cal L}_{\rm (tail)^2+tail(tail)}$ and ${\cal L}_{\rm inst}$
should cancel out, so that the physical energy flux ${\cal L}$ is indeed
independent of $r_0$.  We leave for future work [56,57] the check of the
latter assertion.

\appendix{Formulas to compute the cubic non-linearities} 

In this Appendix we compute the finite part of the retarded integral of
a source term with multipolarity $\ell$, radial dependence with $k \geq 1$,
and containing a non-local integral whose kernel is a certain Legendre
function $Q_m(x)$ [see (2.9)].  Thus,

$$_{k,m}\Psi_L = {\rm FP}_{B=0} \square^{-1}_R \left[ (r/r_0)^B
\hat{n}_L r^{-k} \int^{+\infty}_1 dx Q_m (x) F (t - rx) \right] \ .\eqno
({\rm A}.1) $$

We tackle first the case $k=1$.  It can be checked in this case that the
source behaves when $r \to 0$ is such a way that the retarded integral
(in its usual triple integral form) is convergent, so we can forget
about the finite part procedure.  For the present computation it is
convenient to use the particular formula (D5) of Appendix D in [16],
which gives

$$\eqalignno{
_{1,m}\Psi_L = - {\hat{n}_L\over 2r} \int^{t - r}_{-\infty} d \xi
\int^{t + r - \xi\over 2}_{t - r - \xi\over 2} d &w \int^{+\infty}_1
dx Q_m (x) F (\xi - (x-1) w) \cr
&\times P_\ell \left( 1 - {\left( t - r - \xi \right) \left( t +
r - \xi - 2w \right)\over 2rw} \right) \ ,\qquad &({\rm A}.2)  \cr  }$$
where $P_\ell$ denotes the Legendre polynomial.  Changing the variables
$(\xi, w)$ to the new variables $(y, z)$ defined by $\xi - (x-1) w =
t-ry$ and $z = 1 - ( t - r - \xi) ( t + r - \xi - 2w )/2rw$, we obtain

$$\eqalignno{
_{1,m}\Psi_L = - {r \hat{n}_L\over 2} \int^{+\infty}_1 {dx\over {x^2 -
1}} Q_m&(x) \int^{+\infty}_1 dy F (t - ry) \int^1_{-1} dz~P_\ell (z) \cr
&\times \left[ {xy - z\over \sqrt{(xy - z)^2 - (x^2 - 1) (y^2 -
1)}} - 1 \right] \ . \qquad &({\rm A}.3) \cr  }$$
Next we integrate by parts the $y$-integral, introducing the
anti-derivative $F^{(-1)}$ of $F$.  After some manipulations we get

$$\eqalignno{
_{1,m}\Psi_L = {\hat{n}_L\over 2} \int^{+\infty}_1 dy F^{(-1)} (&t - ry)
\int^{+\infty}_1 dx Q_m (x) \cr
&\times {d\over dx} \left[ \int^1_{-1} {dz~P_\ell (z)\over \sqrt{(xy - z)^2
- (x^2 - 1) (y^2 - 1)}} \right] \ . \qquad &({\rm A}.4) \cr  }$$
The $z$-integration can be performed explicitly thanks to the rather
interesting mathematical formula

$${1\over 2} \int^1_{-1} {dz P_\ell (z)\over \sqrt{(xy - z)^2 - (x^2 - 1)
(y^2 - 1)}} = \left\{\matrix{ P_\ell (x) Q_\ell (y)  & & ~(1 < x \leq
y) \ , \cr P_\ell (y) Q_\ell (x)  & & ~(1 < y \leq x)\ . \cr } \right.
\eqno ({\rm A}.5)  $$
As this formula does not seem to appear in standard text-books of mathematical
formulas such as [47] we present its proof at the end of this Appendix.

With (A.5) we obtain

$$\eqalignno{
_{1,m}\Psi_L = \hat{n}_L \int^{+\infty}_1 dy F^{(-1)} (t - ry&)  
 \biggl\{ Q_\ell (y) \int^y_1 dx Q_m (x) {dP_\ell\over dx} (x) \cr
 &+ P_\ell (y) \int^{+\infty}_y dx Q_m (x)
  {dQ_\ell\over dx} (x) \biggr\} \ . \qquad &({\rm A}.6) \cr }$$
The leading-order term at infinity ($r \to \infty$, $t-r=$ const) comes from
the second term in the brackets, and we have

$$\eqalignno{
_{1,m}\Psi_L &= {\hat{n}_L\over r} \int^{+\infty}_0 d \tau F^{(-1)}
(t-r - \tau)
\int^{+\infty}_{1+\tau /r} dx Q_m (x) {d Q_\ell\over dx}
(x) + o \left( {r^{\varepsilon-2}} \right) \qquad &({\rm A}.7) \cr  }$$
(see the notation concerning the remainder in Section 3.2). 
In the case $\ell=m$ (the only one needed in this paper), the
$x$-integral is obtained in closed-form, and we obtain

$$_{1,\ell}\Psi_L = - {\hat{n}_L\over 2r} \int^{+\infty}_0 d \tau F^{(-1)}
(t-r - \tau) \left[ Q_\ell \left(1 +
{\tau \over r}\right) \right]^2 + o \left( {r^{\varepsilon-2}}
\right) \ . \eqno ({\rm A}.8)  $$
Using the expansion (2.9c) of the Legendre function we further obtain

$$\eqalignno{
 _{1,\ell}\Psi_L = &-{\hat{n}_L\over 8r} \int^{+\infty}_0 d\tau F^{(-1)}
  (t-r - \tau) \cr
&\times \biggl[ \ln^2 \left( {\tau\over 2r} \right) + 4 \biggl(
  \sum^\ell_{j=1} {1\over j} \biggr) \ln \left( {\tau\over 2r} \right) + 4
  \biggl( \sum^\ell_{j=1} {1\over j} \biggr)^2 ~\biggr]
  + o\left(r^{\varepsilon-2}\right)  \ . \quad &({\rm A}.9) \cr  }$$
This is the formula used in Section 3.2. 

Consider now the case $k \geq 2$, and restrict attention to the
leading-order term at infinity.  In this case one uses Lemma 7.2 in
[16], whose hypothesis are satisfied with $N=k-\varepsilon \geq
2-\varepsilon$.  Thus the $1/r$ term in $_{k,m}\Psi_L$, which is deduced from
(7.2) and (7.6) in [16], is

$$_{k,m}\Psi_L = {(-)^\ell\over \ell !} {\hat{n}_L\over r} G^{(\ell)} (u)
  + O \left( {r^{\varepsilon-2}} \right) \ , \eqno ({\rm A}.10) $$
where $u=t-r$ and

$$\eqalignno{
  G (u) = {\rm FP}_{B=0} \int^{u}_{-\infty} &ds \ R_{+\infty}^B
  \left({u - s\over 2} , s \right) \ ,    &({\rm A}.11a) \cr
  R_{+\infty}^B (\rho , s) = - \rho^\ell \int^{+\infty}_\rho &dy {(\rho -
  y)^\ell\over \ell !} \left( {2\over y} \right)^{\ell - 1}
  \left( {y\over r_0}\right)^B y^{-k} \cr
  \times\int^{+\infty}_1 &dx Q_m (x) F(s - (x-1)y)\ . &({\rm A}.11b) \cr}$$
By combining together these expressions, and introducing the variable
$z=1-(u-s)/y$, we get

$$\eqalignno{
G(u) = - {1\over 2^{\ell +1}} {\rm FP}_{B=0} \int^{+\infty}_1 &dx Q_m (x)
\int^{+\infty}_0 dy \left( {y\over r_0} \right)^B y^{-k+\ell +2} \cr
\times \int^1_{-1} &dz~(z^2 - 1)^\ell F (u - (x-z)y)\ .&({\rm A}.12) \cr }$$
Still this expression is not suitable for practical computations, and we
must integrate by parts the $y$-integral in order to determine the
finite part at $B=0$.  The all-integrated terms during the integrations
by parts are zero by analytic continuation in $B$.  We must distinguish
two cases, $2 \leq k \leq \ell +2$ and $k \geq \ell +3$.

When $2 \leq k \leq \ell +2$ we obtain after $\ell+2-k$ integrations by
parts the local expression

$${(-)^\ell\over \ell !}~G^{(\ell )} (u) = - \
_{k,m}\alpha_\ell ~F^{(k-3)} (u) \ ,  \eqno ({\rm A}.13)  $$
where the coefficient $_{k,m}\alpha_\ell$ reads

$$_{k,m}\alpha_\ell = {(\ell - k + 2)!\over 2^{\ell +1} \ell !}
\int^{+\infty}_1 dx Q_m (x) \int^1_{-1} dz~{(1-z^2)^\ell\over
(x-z)^{\ell - k + 3}}  \ . \eqno ({\rm A}.14)  $$
A more elegant form of this coefficient is obtained by using the
representation of the Legendre function given by 
   
$$ Q_\ell (x) = {1\over 2^{\ell +1}} \int^1_{-1} dz~{(1-z^2)^\ell\over
  (x-z)^{\ell +1}}     \eqno ({\rm A}.15)  $$
(which is equivalent to the other formulas (2.9), see e.g.  [52]).  Then
$_{k,m}\alpha_\ell$ reads

$$_{k,m}\alpha_\ell = \int^{+\infty}_1 dx Q_m (x) \int^{+\infty}_x dz
{(z-x)^{k-3}\over (k-3)!} Q_\ell (z)  \ .  \eqno ({\rm A}.16)   $$
The numerical values of the coefficient are computed from

$$\eqalignno{
_{k,m}\alpha_\ell = \sum^{k-2}_{i=0} &\ {(-)^i(k-2)! \over {i! (k-i-2)!}}
{(2\ell - 2k +3 +2i)!!\over (2\ell +1+2i)!!} \cr
&\times (2\ell -2k + 5 + 4i) 
\int^{+\infty}_1 dx Q_m (x) Q_{\ell -k+2+2i}(x) \ , 
&({\rm A}.17)  \cr   }$$
where the remaining integrals are given by (see e.g. [47])

$$
\int^{+\infty}_1 dx Q_m (x) Q_p (x) =  \left\{ \matrix{ {1\over (m-p )
(m+p+1)} \left[ \sum^m_{j=1} {1\over j} - \sum^p_{j=1} {1\over j}
\right]  \quad & (m \not = p )\ , \cr
{1\over 2p +1} \left[ {\pi^2\over 6} - \sum^p_{j=1} {1\over j^2}
\right]  \hfill \quad & (m=p )\ . \cr } \right.  \eqno ({\rm A}.18)  $$
We thus obtain the coefficients reported in Tables 1 and 2 of Section 3.

In the case $k \geq \ell +3$ we get after $k - \ell -2$ integrations by
parts the non-local expression

$${(-)^\ell\over \ell !} G^{(\ell)} (u) = - \int^{+\infty}_0 d \tau
F^{(k-2)} (u - \tau) \left[ _{k,m}\beta_\ell \ln \left( {\tau\over
2r_0} \right) + \ _{k,m}\gamma_\ell \right] \ , \eqno({\rm A}.19)
$$
where the coefficients $_{k,m}\beta_\ell$ and $_{k,m}\gamma_\ell$ are
given by

$$\eqalignno{
_{k,m}\beta_\ell = {1\over 2^{\ell +1} \ell ! (k-\ell -3)!}
\int^{+\infty}_1 dx Q_m (x) &\int^1_{-1} dz (1-z^2)^\ell (z-x)^{k-\ell
-3}  \ , \qquad\quad &({\rm A}.20a) \cr
_{k,m}\gamma_\ell = {1\over 2^{\ell +1} \ell ! (k-\ell -3)!}
\int^{+\infty}_1 dx Q_m (x) &\int^1_{-1} dz (1-z^2)^\ell
(z-x)^{k-\ell -3} \cr &\times \biggl[ -\ln \left( {x-z\over
2} \right) + \sum^{k-\ell -3}_{j=1} {1\over j} \biggr] \ . &({\rm A}.20b)}$$
More elegant forms are

$$\eqalignno{
_{k,m}\beta_\ell = {1\over 2} \int^{+\infty}_1 dx Q_m (x) &\int^1_{-1}
dz {(z-x)^{k-3}\over (k-3)!} P_\ell (z) \ , &({\rm A}.21a)     \cr
_{k,m}\gamma_\ell = {1\over 2} \int^{+\infty}_1 dx Q_m (x) &\int^1_{-1}
dz {(z-x)^{k-3}\over (k-3)!} P_\ell (z) 
\biggl[ -\ln \left( {x-z\over 2} \right) +
\sum^{k-3}_{j=1} {1\over j} \biggr] \ .  \cr & &({\rm A}.21b)   }$$
The cases $k=3$ and $\ell=0$ needed in this paper yield

$$\eqalignno{
_{3,m}\beta_0 &= \int^{+\infty}_1 dx Q_m (x) = {1\over m(m+1)}
\qquad  (m \geq 1) \ ,   &({\rm A}.22a)  \cr
_{3,m}\gamma_0 &= - \int^{+\infty}_1 dx Q_m (x) \biggl[ \ln \left(
{1\over 2}\sqrt{x^2 - 1} \right) + Q_1 (x)  \biggr] \ . &({\rm A}.22b)\cr} $$
Finally the formula 7.132.3 in [47] where, however, a factor
$2^{2\lambda-\mu}$ in the denominator should be corrected to a factor
$2^{2-\mu}$, leads to

$$_{3,m}\gamma_0 = \left\{ \matrix{ - {\pi^2\over 18} + {7\over 12} 
\hfill \quad & (m=1)\ , \cr
{1\over m(m+1)(m+2)} \left( {m^2 +3m +3\over m+1} - {2\over m-1}
\sum^{m-1}_{j=1} {1\over j} \right)  \quad & (m \geq 2) \ .\cr }  \right.
\eqno ({\rm A}.22c)   $$
When $m=2$ we find the values (3.13).

{\it Proof of the mathematical formula ({\rm A.5})}.
The author proved this formula by (i) verifying that the left-hand-side
of the formula, namely

$$I_\ell(x,y) \equiv {1\over 2} \int^1_{-1}
{dz P_\ell (z)\over \sqrt{(xy - z)^2 - (x^2 - 1)(y^2 - 1)}} \ ,
\eqno ({\rm A}.23)   $$
is a particular solution of the Legendre equation both in variables $x$
and $y$, and (ii) showing that this particular solution $I_\ell(x,y)$ is
necessarily equal to the solution given by the right-hand-side.  To
prove (ii) one must invoke the behaviour of the Legendre function at
infinity, which is $Q_\ell(x) \approx 1/x^{\ell+1}$ when $x \to
\infty$.

Then a direct and more interesting proof of this formula was found by H. Sivak.
This proof consists first of transforming $I_\ell(x,y)$ by means of the
formula 3.613.1 of Gradshteyn and Ryzhik [47] in order to obtain

$$I_\ell(x,y) = {1\over 2\pi} \int^\pi_0 dt \int^1_{-1}
{dz\,P_\ell (z)\over {xy-z-\sqrt{(x^2 - 1)(y^2 - 1)} ~{\rm cos}t}} \ .
\eqno ({\rm A}.24)   $$
With the help of Neumann's formula (2.9a) for the Legendre function we get

$$I_\ell(x,y) = {1\over \pi} \int^\pi_0 dt ~Q_\ell
(xy-\sqrt{(x^2 - 1)(y^2 - 1)} ~{\rm cos}t) \ . \eqno ({\rm A}.25)   $$
Using now the formula 8.795.2 of [47] we find (assuming for instance that
$x<y$)

$$I_\ell(x,y) = {1\over \pi} \int^\pi_0 dt \biggl\{P_\ell(x)Q_\ell (y)
+2\sum_{k=1}^{+\infty}(-)^k P_\ell^{-k}(x)Q_\ell^k(y)~{\rm cos}kt\biggr\} \ .
\eqno ({\rm A}.26)   $$
This yields immediately the desired result

$$I_\ell(x,y) = P_\ell(x)Q_\ell (y)\ .  \eqno ({\rm A}.27)   $$

\appendix{The harmonicity algorithm in the far zone} 

The harmonicity algorithm is a mean to add to the finite part of the
retarded integral of the source term,

$$u^{\alpha\beta} = {\rm FP}_{B=0} \square^{-1}_R \left[
  \left( {r\over r_0}\right)^B \Lambda^{\alpha\beta} \right] \ ,
  \eqno({\rm B}.1) $$
an homogeneous solution $v^{\alpha\beta}$ of the wave equation which is
such that the sum $u^{\alpha\beta}+v^{\alpha\beta}$ is divergence-free
(and thus satisfies the harmonic-gauge condition).  A particular
algorithm is defined by (4.12)-(4.13) in [16];  a slightly different one
by (2.11)-(2.12) in paper I.  Here we follow paper I.

The harmonicity algorithm to order $1/r$ when $r \to \infty$, $t-r=$
const is as follows.  The $1/r$ term of the divergence of
$u^{\alpha\beta}$,

$$\partial_\beta u^{\alpha\beta} = {1\over r} U^\alpha ({\bf n}, t-r)
+ O \left( {1\over r^2} \right) \ ,    \eqno ({\rm B}.2)   $$
is decomposed into multipole moments according to

$$ \eqalignno{
 U^0 &= \sum_{\ell \geq 0} n_L W_L  \ ,  &({\rm B}.3a) \cr
 U^i &= \sum_{\ell \geq 0} n_{iL} X_L
 + \sum_{\ell \geq 1} \bigl\{ n_{L-1} Y_{iL-1} + \varepsilon_{iab}
 n_{aL-1} Z_{bL-1}  \bigr\} \ .  &({\rm B}.3b)  \cr   }$$
Comparing this decomposition with the definition (2.11) in paper I, we
obtain some relations between the tensors $W_L$, $X_L$, etc., and the
time-derivatives of the tensors $A_L$, $B_L$, etc., in this definition.
Next we obtain the $1/r$ term of $v^{\alpha\beta}$,

$$v^{\alpha\beta} = {1\over r} V^{\alpha\beta} ({\bf n}, t-r) + O
\left( {1\over r^2} \right) \ ,  \eqno ({\rm B}.4)  $$
by applying the formula (2.12) in paper I.  Re-expressing the
decomposition in terms of the tensors $W_L$, $X_L$, etc., we obtain some
formulas for the time-derivative of $V^{\alpha\beta}$,

$$   \eqalignno{
{dV^{00}\over dt} &=-W -n_a \left[ W_a +Y_a -3 X_a \right]\ ,&({\rm B}.5a)\cr
 {dV^{0i}\over dt} &=- Y_i + 3 X_i - \varepsilon_{iab} n_a Z_b +
   \sum_{\ell \geq 2} n_{L-1} W_{iL-1}  \ ,  &({\rm B}.5b) \cr
 {dV^{ij}\over dt} &= \delta_{ij} X + \sum_{\ell \geq 2} \biggl\{
   -2 \delta_{ij} n_{L-1} X_{L-1} + 6 n_{L-2(i} X_{j)L-2} \cr
 &\qquad\qquad\quad\quad
+n_{L-2} \left[ W_{ijL-2} -3 X_{ijL-2} +Y_{ijL-2} \right] \cr
 &\qquad\qquad\quad\quad +2 n_{aL-2} \varepsilon_{ab(i} Z_{j)bL-2} \biggr\} \ . 
 &({\rm B}.5c)    \cr } $$
These are the formulas needed in Section 4.1.

\ack
The author would like to thank Horacio Sivak for his proof of the mathematical
formula (A.5) and his permission of presenting it at the end of Appendix A.

\references
\numrefjl{[1]}{Blanchet L}{Quadrupole-quadrupole gravitational
waves}{}{(referred to as paper I)}
\numrefjl{[2]}{Four\`es-Bruhat F 1952}{Acta Math.}{88}{141}
\numrefjl{[3]}{Bruhat Y 1964}{Ann. Math. Pura Appl.}{64}{191}
\numrefjl{[4]}{Kundt W and Newman E T 1968}{J. Math. Phys.}{9}{2193}
\numrefjl{[5]}{McLenaghan R G 1969}{Proc. Camb. Philos. Soc.}{65}{139}
\numrefbk{[6]}{Friedlander F G 1975}{The wave equation on a curved
space-time}{(Cambridge: Cambridge University Press)}
\numrefjl{[7]}{Waylen P C 1978}{Proc. R. Soc. London A}{362}{233}
\numrefjl{[8]}{Choquet-Bruhat F, Christodoulou D and
Francaviglia M 1979}{Ann. Inst. Henri Poincar\'e A}{31}{399}
\numrefjl{[9]}{Carminati J and McLenaghan R G 1986}{Ann. Inst.
Henri Poincar\'e (Physique Th\'eorique) A}{44}{115}
\numrefjl{[10]}{Bonnor W B 1959}{Philos. Trans. R. Soc.
London A}{251}{233}
\numrefjl{[11]}{Bertotti B and Plebanski J 1960}{Ann. Phys.
(N.Y.)}{11}{169}
\numrefjl{[12]}{Thorne K S and Kov\'acs S J 1975}{Astrophys.
J.}{200}{245}
\numrefjl{[13]}{Crowley R J and Thorne K S 1977}{Astrophys.
J.}{215}{624}
\numrefjl{[14]}{Thorne K S 1980}{Rev. Mod. Phys.}{52}{299}
\numrefjl{[15]}{Bel L, Damour T, Deruelle N, Iba\~nez J and
Martin J 1981}{Gen. Relativ. Gravit.}{13}{963}
\numrefjl{[16]}{Blanchet L and Damour T 1986}{Philos. Trans. R. Soc.
London A}{320}{379}
\numrefjl{[17]}{Blanchet L 1987}{Proc. R. Soc. Lond. A}{409}{383}
\numrefjl{[18]}{Peters P C 1966}{Phys. Rev.}{146}{938}
\numrefjl{[19]}{Price R H 1972}{Phys. Rev. D}{5}{2419}
\numrefjl{[20]}{Price R H 1972}{Phys. Rev. D}{5}{2439}
\numrefjl{[21]}{Bardeen J M and Press W H 1973}{J. Math. Phys.}{14}{7}
\numrefjl{[22]}{Schmidt B G  and Stewart J M 1979}{Proc. R. Soc. London
A}{367}{503}
\numrefjl{[23]}{Porrill J and Stewart J M 1981}{Proc. R. Soc. London
A}{376}{451}
\numrefjl{[24]}{Leaver E W 1986}{Phys. Rev.}{34}{384}
\numrefjl{[25]}{Bonnor W B and Rotenberg M A 1966}{Proc. R. Soc. London
A}{289}{247}
\numrefjl{[26]}{Couch W E, Torrence R J, Janis A I and
Newman E T 1968}{J.  Math.  Phys.}{9}{484}
\numrefjl{[27]}{Hunter  A J and Rotenberg M A 1969}{J. Phys. A}{2}{34}
\numrefbk{[28]}{Bonnor W B  1974 in}{Ondes et radiations
gravitationnelles}{(CNRS, Paris, 1974) p. 73}
\numrefjl{[29]}{Blanchet L and Damour T 1988}{Phys. Rev. D}{37}{1410}
\numrefjl{[30]}{Blanchet L and Sch\"afer G 1993}{Class.
Quantum Grav.}{10}{2699}
\numrefjl{[31]}{Blanchet L and Sathyaprakash B S 1995}{Phys. Rev.
Letters}{74}{1067}
\numrefjl{[32]}{Blanchet L and Damour T 1992}{Phys. Rev. D}{46}{4304}
\numrefjl{[33]}{Blanchet L 1995}{Phys. Rev. D}{51}{2559}
\numrefjl{[34]}{Will C M and Wiseman A G 1996}{Phys. Rev. D}{54}{4813}
\numrefjl{[35]}{Blanchet L 1996}{Phys. Rev. D}{54}{1417}
\numrefjl{[36]}{Cutler C, Apostolatos T A, Bildsten L, Finn L S,
Flanagan E E, Kennefick D, Markovic D M, Ori A, Poisson E,
Sussman  G J and Thorne K S 1993}{Phys. Rev. Lett.}{70}{2984}
\numrefjl{[37]}{Cutler C, Finn L S, Poisson E and Sussmann G J
1993}{Phys.  Rev.  D}{47}{1511}
\numrefjl{[38]}{Tanaka T, Shibata M, Sasaki M, Tagoshi H
and Nakamura T 1993}{Prog. Theor. Phys.}{90}{65}
\numrefjl{[39]}{Tagoshi H and Nakamura T 1994}{Phys. Rev.
D}{49}{4016}
\numrefjl{[40]}{Poisson E 1995}{Phys. Rev. D}{52}{5719}
\numrefjl{[41]}{Our notation is the following:  signature $-+++$;  greek
indices =0,1,2,3;  latin indices =1,2,3;  $g={\rm det}$ $(g_{\mu\nu})$;
$\eta_{\mu\nu}= \eta^{\mu\nu}$ =~flat metric =~diag (-1,1,1,1);
$r=|{\bf x}|=(x_1^2 +x_2^2 +x_3^2)^{1/2}$;  $n^i =n_i =x^i/r$;
$\partial_i =\partial/\partial x^i$; $n^L =n_L= n_{i_1} n_{i_2}\ldots
n_{i_\ell}$ and $\partial_L =\partial_{i_1} \partial_{i_2}\ldots
\partial_{i_\ell}$, where $L=i_1 i_2\ldots i_\ell$ is a multi-index with
$\ell$ indices; $n_{L-1} = n_{i_1} \ldots n_{i_{\ell-1}}$, $n_{aL-1}
=n_a n_{L-1}$, etc\dots; $\hat n_L$ and $\hat\partial_L$ are the (symmetric)
and trace-free (STF) parts of $n_L$ and $\partial_L$, also denoted by
$n_{<L>}$, $\partial_{<L>}$;  the superscript $(n)$ denotes $n$ time
derivatives;  $T_{(\alpha\beta)} ={1\over 2} (T_{\alpha\beta} +
T_{\beta\alpha})$ and $T_{(ij)} = {1\over 2} (T_{ij} + T_{ji})$}{}{}{}
\numrefbk{[42]}{Fock V A 1959}{Theory of Space, Time and
Gravitation}{Pergamon, London}
\numrefjl{[43]}{Madore J 1970}{Ann. Inst. H. Poincar\'e}{12}{285
and 365}
\numrefjl{[44]}{Anderson J L, Kates R E, Kegeles L S
and Madonna R G 1982}{Phys. Rev. D}{25}{2038}
\numrefjl{[45]}{Blanchet L, Damour T and Iyer B R
1995}{Phys. Rev. D}{51}{5360}
\numrefbk{[47]}{Gradshteyn I S and Ryzhik I M 1980}{Table of Integrals,
Series and Products}{(Academic Press, Inc.)}
\numrefjl{[48]}{Poisson E 1993}{Phys. Rev. D}{47}{1497}
\numrefjl{[49]}{Sasaki M 1994}{Prog. Theor.
Phys.}{92}{17}
\numrefjl{[50]}{Tagoshi H and Sasaki M 1994}{Prog. Theor.
Phys.}{92}{745}
\numrefjl{[51]}{Tanaka T, Tagoshi H and Sasaki M 1996}{Prog. Theor.
Phys.}{96}{1087}
\numrefbk{[52]}{Whittaker E T and Watson G N 1990}{A course of
Modern Analysis}{reprinted edition (Cambridge U. Press)}
\numrefjl{[53]}{Leonard S W and Poisson E}{Phys. Rev. D}{}{in press
gr-qc/9705014}
\numrefjl{[54]}{Blanchet L and Damour T 1989}{Ann. Inst. H.
Poincar\'e}{50}{377}
\numrefjl{[55]}{Damour T and Iyer B R 1991}{Ann. Inst. H.
Poincar\'e}{54}{155}
\numrefjl{[56]}{Blanchet L, Iyer B R and Joguet B}{}{}{in preparation}
\numrefjl{[57]}{Blanchet L, Faye G, Iyer B R, Joguet B and Ponsot B}{}{}{in
progress}
\numrefjl{[58]}{Damour T, Iyer B R and Sathyaprakash B S}{}{}{gr-qc/9708034}
\bye